\documentclass[11pt,preprint]{aastex}
\usepackage{epsfig}
\usepackage{natbib}
\usepackage{graphicx}
\usepackage{slashbox}
\usepackage{multirow}
\usepackage{lscape}
\usepackage{mathrsfs,amssymb}
\usepackage{amsmath}

\newcommand       \K            {\,{\rm K}}

\newcommand       \mum          {\,{\rm \mu m}}

\newcommand       \Teff         {T_{\rm eff}}

\newcommand       \Lstar        {L_\star}

\newcommand       \simali       {\sim\,}

\def    \W      {\,{\rm W}}
\def    \m      {\,{\rm m}}
\def    \Mdust  {M_{\rm dust}}

\def    \Ftwenty     {F_{21}}
\def    \Fthirty     {F_{30}}
\def    \Fuir     {F_{\rm UIR}}
\def    \Fir      {F_{\rm IR}}
\def    \Mtwenty     {M_{\rm 21\mu m}}
\def    \Mthirty     {M_{\rm 30\mu m}}
\def    \Muir        {M_{\rm UIR}}
\def    \kappatwenty     {\kappa_{\rm abs}^{\rm 21\mu m}}
\def    \kappathirty     {\kappa_{\rm abs}^{\rm 30\mu m}}
\def    \kappauir        {\kappa_{\rm abs}^{\rm UIR}}
\def    \fractwenty     {f_{\rm 21}}
\def    \fracthirty     {f_{\rm 30}}
\def    \fracuir        {f_{\rm UIR}}

\countdef\decade=200
\advance\decade by \year
\countdef\hours=201
\advance\hours by \time
\divide\hours by 60
\countdef\mins=202
\advance\mins by \hours
\multiply\mins by 60
\multiply\hours by 100
\countdef\miltime=203
\advance\miltime by \hours
\advance\miltime by \time
\advance\miltime by -\mins
\def\today{\number\decade.\number\month.\number\day.\number\miltime}

\shorttitle{On the Mysterious 21$\mum$, 30$\mum$ and UIE Features}
\title{
\vspace*{-2.0em}
{\normalsize\rm Accepted for publication in
           {\it The Astrophysical Journal}}\\
\vspace*{1.0em}
A Tale of Three Mysterious Spectral Features 
in Carbon-Rich Evolved Stars: 
The 21$\mum$, 30$\mum$, and 
``Unidentified Infrared'' Emission Features
\\{\small DRAFT: \today ~~}
}
\author{Ajay Mishra\altaffilmark{1},
        Aigen Li\altaffilmark{1},
        and B.W.~Jiang\altaffilmark{2}}
\altaffiltext{1} {Department of Physics and Astronomy,
                  University of Missouri,
                  Columbia, MO 65211, USA;
                  {\sf amishra@mail.missouri.edu,
                       lia@missouri.edu}
                   }
\altaffiltext{2}{Department of Astronomy,
                 Beijing Normal University,
                 Beijing 100875, China;
                 {\sf bjiang@bnu.edu.cn}
                 }

\begin{document}

\begin{abstract}
The mysterious ``21$\mum$'' emission feature 
seen almost exclusively in the short-lived 
protoplanetary nebula (PPN) phase of stellar 
evolution remains unidentified since its discovery 
two decades ago. This feature is always accompanied 
by the equally mysterious, unidentified ``30$\mum$'' 
feature and the so-called ``unidentified infrared'' 
(UIR) features 
at 3.3, 6.2, 7.7, 8.6, and 11.3$\mum$ which are
generally attributed to polycyclic 
aromatic hydrocarbon (PAH) molecules. 
The 30$\mum$ feature is commonly observed
in all stages of stellar evolution from 
the asymptotic giant branch (AGB) through PPN 
to the planetary nebula phase. 
We explore the interrelations among the mysterious 
21$\mum$, 30$\mum$, and UIR features 
of the 21$\mum$ sources. 
We derive the fluxes emitted in the observed UIR, 
21$\mum$, and 30$\mum$ features from 
published {\it ISO} or {\it Spitzer}/IRS spectra. 
We find that none of these spectral features 
correlate with each other. 
This argues against a common carrier 
(e.g., thiourea) for both the 21$\mum$ 
feature and the 30$\mum$ feature.
This also does not support large PAH clusters 
as a possible carrier for the 21$\mum$ feature.
\end{abstract}
\keywords{circumstellar matter --- dust, extinction 
          --- infrared: stars --- stars: AGB and Post-AGB 
          --- stars: evolution}

\section{Introduction\label{sec:intro}}
In carbon-rich evolved objects, there are two prominent, 
mysterious emission bands known as the ``21$\mum$'' 
and ``30$\mum$'' spectral features (see Jiang et al.\ 2010).
The enigmatic 21$\mum$ emission feature was first detected in
four protoplanetary nebulae (PPNe) through the 7.7--22.6$\mum$ 
spectra obtained by the {\it Low Resolution Spectrometer} (LRS)
on board the {\it Infrared Astronomical Satellite} 
(IRAS; Kwok et al.\ 1989).
To date, this feature has been seen unambiguously 
in $\simali$18 objects in the Milky Way (Cerrigone et al.\ 2011)
and $\simali$9 objects in the Large and Small
Magellanic Clouds (Volk et al.\ 2011).\footnote{%
   In the Magellanic Clouds the total number of
   21$\mum$ sources would be 13 if the marginal detection
   in the other four sources is included (Volk et al.\ 2011).
   }
The 21$\mum$ sources are all PPNe,
objects in a rapid transition period 
(of several thousand years) between 
the asymptotic giant branch (AGB)
and planetary nebula (PN) phases.\footnote{%
   A few weak detections of the 21$\mum$ feature have also been   
   reported for their precursors 
   -- highly evolved carbon stars (Volk et al.\ 2000)
   and their successors -- PNe (Hony et al.\ 2001, Volk 2003).
   These detections are yet not very conclusive as the 21$\mum$
   feature claimed for these sources is considerably weak.
   }
With a peak wavelength at $\simali$20.1$\mum$ 
and a FWHM of $\simali$2.2--2.3$\mum$,
the 21$\mum$ feature displays little shape variation
among different sources.
This feature emits up to $\sim$\,8\% of the total
infrared (IR) power of a 21$\mum$ source.
%
The carrier of this feature remains unidentified,
despite the fact that over a dozen carrier
candidates have been proposed,
including inorganic materials: 
TiC (von Helden et al.\ 2000, but see Li 2003),
SiS$_2$ (Goebel 1993),
SiC (Speck \& Hofmeister 2004, but see Jiang et al.\ 2005),
SiO$_2$-coated SiC (Posch et al.\ 2004),
carbon-silicon mixtures (Kimura et al.\ 2005), 
FeO (Posch et al.\ 2004, but see Li et al.\ 2013), 
Fe$_2$O$_3$ and Fe$_3$O$_4$ 
(Cox 1990, but see Zhang et al.\ 2009a),
and organic materials: 
Ti-coordinated caged carbon (Kimura et al.\ 2005), 
urea or thiourea (Sourisseau et al.\ 1992), 
polycyclic aromatic hydrocarbon (PAH), 
and hydrogenated amorphous carbon (HAC) 
(Buss et al.\ 1993, Justtanont et al.\ 1996). 
It is considered as ``{\it one of the most 
interesting unresolved mysteries 
in astrochemistry}'' (Kwok et al.\ 2002).

The 30$\mum$ feature, first discovered by 
Forrest et al.\ (1981) in several carbon stars 
and in two PNe through 
the {\it Kuiper Airborne Observatory} spectrometry,
is very broad and strong,
extending from $\simali$24$\mum$ to $\simali$45$\mum$ and 
accounting for up to $\simali$30\% of the total IR luminosity 
of a 30$\mum$ source (Volk et al.\ 2002).
Unlike the 21$\mum$ feature which displays little
shape variation, the 30$\mum$ feature varies in its 
peak wavelength and width among different sources
(e.g., see Hrivnak et al.\ 2000, 
Hony et al.\ 2002). 
%

The carrier of the 30$\mum$ feature also remains unidentified.
Magnesium sulfide (MgS) solids have long been proposed as 
a candidate carrier 
(Goebel \& Moseley 1985; Nuth et al.\ 1985;
Jiang et al.\ 1999; Szczerba et al.\ 1999; 
Hony et al.\ 2002, 2003;
Lombaert et al.\ 2012).
However, Zhang et al.\ (2009b) recently 
ruled out MgS as a valid carrier since it 
would require too much S to account for
the observed large amount of power emitted 
from the 30$\mum$ feature
(also see Messenger et al.\ 2013,
Otsuka et al.\ 2014).\footnote{%
  Lombaert et al.\ (2012) modeled the 30$\mum$ feature 
  of the extreme carbon star LL Peg with MgS dust. 
  They argued that if MgS is in thermal contact 
  with amorphous carbon and SiC, the amount of MgS 
  required to reproduce the strength of the 30$\mum$ feature 
  agrees with the solar abundance of sulfur.
  }

While so far the 21$\mum$ feature appears to be restricted 
to PPNe, the 30$\mum$ feature is more ubiquitously seen in 
a large number of carbon-rich objects 
at various evolutionary stages,
including AGB stars, post-AGB stars and PNe
(see Jiang et al.\ 2010, Zhang \& Jiang 2008).
We note that {\it the 30$\mum$ feature is seen in all 21$\mum$ sources,}
although not all 30$\mum$ sources emit at the 21$\mum$ feature. 
Furthermore, all 21$\mum$ sources display the so-called
``unidentified IR'' (UIR) features 
at 3.3, 6.2, 7.7, 8.6 and 11.3$\mum$ 
(Hrivnak et al.\ 2008) which are commonly 
attributed to PAH molecules 
(L\'eger \& Puget 1984, Allamandola et al.\ 1985).

To gain insight into the nature of the carriers of
the 21$\mum$ and 30$\mum$ features, 
we select ten well-studied Galactic 21$\mum$ sources.
The 21$\mum$, 30$\mum$ and UIR features of these 
sources have been observationally obtained. 
This allows us to derive the fractional fluxes 
emitted respectively in the 21$\mum$, 30$\mum$ 
and UIR features (see \S\ref{sec:fluxes}).
We examine in \S\ref{sec:2130uir} the interrelations 
among the 21$\mum$, 30$\mum$ and UIR features in 
the selected 21$\mum$ sources. 
We discuss in \S\ref{sec:implications} 
the implications of these interrelations.
%
\S\ref{sec:continuum} discusses 
the methods of deriving the fluxes of 
the 21$\mum$, 30$\mum$ and UIR features. 
%
We summarize the major results in \S\ref{sec:summary}.

%
\section{Deriving the Fluxes of the 21$\mum$, 30$\mum$ 
           and UIR Features\label{sec:fluxes}}
We have selected ten well-studied 
Galactic 21$\mum$ sources.
For these sources, the high-quality mid-IR 
spectra obtained by the {\it Spitzer Space Telescope} 
and the {\it Infrared Space Observatories} (ISO) 
allow one to measure relatively accurately 
the (integrated) fluxes emitted 
in the 21$\mum$ feature ($\Ftwenty$),
in the 30$\mum$ feature ($\Fthirty$),
and in the UIR features ($F_{\rm UIR}$).\footnote{%
  To have a full count of the 21$\mum$, 30$\mum$,
  and UIR fluxes, we require that the IR spectra of 
  the selected sources at least cover the wavelength
  range of $\simali$5--35$\mum$.
  As a result of this criterion, 
  the 21$\mum$ sources which lack 
  the IR spectra at $\lambda<10\mum$  
  are not included in this study.
  }
%
%
In Table~\ref{tab:starpara} 
we tabulate the stellar and circumstellar 
parameters of the ten Galactic 21$\mum$ sources
(compiled from the literature): 
the effective temperature $\Teff$, luminosity $L_\star$,
core mass $M_\star$,
stellar radius $r_\star$, 
and distance $d$ from Earth 
of the central star.
 
\begin{table}
\footnotesize{
\caption{\footnotesize
         \label{tab:starpara}
         Stellar Parameters 
         for the Ten Selected 21$\mum$ Sources.
         }
\begin{center}
\begin{tabular}{cccccccccccc}
\hline
\hline
IRAS & $\Teff$ & $L_{\star}$ & $M_{\star}$ 
     & $r_{\star}$ & $d$ \\ 
\cline{7-10}  
\cline{11-11}
Sources & (K) & ($L_{\odot}$) & ($M_{\odot}$) 
        & ($R_{\odot}$) & (kpc)\\
\hline
02229+6208 & 5500  & 8333  & 0.558   & 98.4  & 2.1  \\ 
04296+3429 & 6500  & 8333  & 0.554   & 69.1  & 5.0  \\
05341+0852 & 6500  & 8430  & 0.551   & 69.5  & 7.8  \\
07134+1005 & 7250  & 6554  & 0.841   & 49.0  & 2.2  \\
16594-4656 & 10000 & 10279 & 0.60    & 32.0  & 2.55 \\
20000+3239 & 5500  & 5186  & ...     & 77.0  & 2.24 \\
22223+4327 & 6500  & 6075  & 0.551   & 59.0  & 3.2  \\
22272+5435 & 5650  & 10990 & 0.574   & 106.0 & 1.55 \\
22574+6609 & 5500  & 8061  & 0.60    & 96.0  & 7.6  \\
23304+6147 & 6750  & 8347  & 0.66    & 64.0  & 3.25 \\
\hline
\end{tabular}
\end{center}
}
\end{table}

Below we briefly comment on the individual sources, 
focusing on the nebular morphology of each source.
{\it IRAS 02229+6208}
is a cool, highly reddened post-AGB star. 
It has an elliptically extended nebula as revealed
by the polarization map of Ueta et al.\ (2005).
{\it IRAS 04296+3429}
is in the advanced post-AGB evolution stage 
with a bipolar lobe structure as revealed 
by the scattered light images
at 0.56 and 0.81$\mum$ (Sahai 1999). 
{\it IRAS 05341+0852}
has a very extended, optically thin atmosphere. 
Its visible image shows an elongated elliptical nebula 
around the central star (Ueta et al.\ 2000).
{\it IRAS 07134+1005 (HD 56126)}
is one of the best studied post-AGB star.
It has an axial symmetric structure 
(Meixner et al.\ 1997; Ueta et al.\ 2000). 
It is one of the sources 
in which the 21$\mum$ and 30$\mum$ features
were first discovered 
(Forrest et al.\ 1981, Kwok et al.\ 1989).
{\it IRAS 16594-4656}
is a bipolar post-AGB star
as indicated by its dust emission 
spectral energy distribution 
(Meixner et al.\ 1999; Hrivnak et al.\ 2008).
It has an optically thick circumstellar envelope.
{\it IRAS 20000+3239} 
has an extended and axisymmetric bipolar structure
as revealed by the near-IR imaging polarimetry of 
this object (Gledhill et al.\ 2001).
For {\it IRAS 22223+4327},
the polarization map reveals an extended, 
optically thick circumstellar envelope
(Gledhill et al.\ 2001). 
{\it IRAS 22272+5435}
is extremely carbon-rich 
and fairly bright both in the IR and optical 
(Ueta et al.\ 2001).
{\it IRAS 22574+6609}
is very faint in the visible. 
It shows a bipolar morphology 
with a dark lane dividing the nebula 
into two lobes (Ueta et al.\ 2000; Su et al.\ 2001). 
{\it IRAS 23304+6147} 
has a quadrupolar shape, although its two pairs 
of lobes are not well separated. This object is 
remarkably similar to IRAS 20000+3239 in its 
post-AGB properties and AGB ejecta 
(Sahai et al.\ 2007).
%




%
We retrieve the IR spectra of the selected sources from 
the {\it Spitzer} and ISO archives. 
All sources exhibit the 21$\mum$, 30$\mum$ and UIR features. 
In order to quantify the relative contributions from different 
components, we perform spectral decomposition of the IR
spectra of these sources.
Following Zhang et al.\ (2010),
we use the IDL package PAHFIT of Smith et al.\ (2007) 
to decompose the observed IR spectra 
into individual components,   
including contributions
from stellar continuum, H$_2$ lines,
thermal dust continuum, 
and the 21$\mum$, 30$\mum$ and UIE features.
This decomposition technique 
will be discussed in \S\ref{sec:continuum}.
We note that the decomposition is not
really a physical dust model. It only serves 
to measure the fluxes in the features.

The stellar continuum is assumed to 
be a blackbody of $T_{\rm eff}$, 
the stellar effective temperature. 
In all  cases, the stellar contribution 
to the IR spectra is negligible.
Following Zhang et al.\ (2010),
we include eight H$_2$ S(0)--S(7) 
rotational lines 
because they are seen in the observed spectra
and need to be subtracted in order to accurately 
determine the fluxes of the dust spectral features.
The profiles of the H$_2$ lines are assumed to be
Gaussian. 
%
A modified blackbody
$I_\lambda\sim\lambda^{-2}\,B_\lambda(T)$
is adopted to simulate the dust thermal continuum,
where $B_\lambda(T)$ is the Planck function
of temperature $T$ at wavelength $\lambda$.
For all the sources
the best-fit requires two components with 
different temperatures to describe the dust 
thermal continuum:
a warm dust component with 
$T_W$\,$\sim$\,130--190$\K$,
and a cold component with 
$T_C$\,$\sim$\,60--90$\K$.
The 21$\mum$, 30$\mum$ and UIR features are 
fitted with a number of Drude profiles 
$I_\lambda =  I_0 \gamma^{2}/
\left\{(\lambda/\lambda_{0}  -  
\lambda_{0}/\lambda)^{2} + \gamma^{2}\right\}$,
where $\lambda_0$ is the central wavelength, 
$I_0$ is the central intensity, 
and $\gamma$ is the fractional FWHM 
of each feature.
The broad 30$\mum$ feature is fitted by two or 
three subfeatures at 26, 30 and 33$\mum$.\footnote{%
   The ISO/SWS spectroscopy suggested that the 30$\mum$ feature 
   consists of two subfeatures: a narrow feature at 26$\mum$ 
   and a broad one at 33$\mum$
   (Hrivnak et al.\ 2000, Volk et al.\ 2002).
   However, the two distinct components in 
   the 30$\mum$ feature are less prominent 
   in the {\it Spitzer} spectra (see Hrivnak et al.\ 2009).
   Like Zhang et al.\ (2010), we find it practical to use 
   two or three Drude profiles peaking at 26, 30 and 30$\mum$ 
   to represent this feature as they provide better fits 
   to the data.
   }
For the UIR features, we include all the PAH
bands listed in Table~1 of Draine \& Li (2007).
The Drude profile is expected for 
classical damped harmonic oscillators
(see Li 2009). It closely resembles 
a Lorentzian profile. Compared with a Gaussian profile,
both the Drude profile and the Lorentzian profile have
more extended wings. The 21$\mum$, 30$\mum$ and UIR features
are better fitted with Drude profiles than Gaussian profiles,
while the H$_2$ lines are better fitted with Gaussian profiles.

\begin{figure*}
\centering
  \begin{tabular}{cc}
    \includegraphics[width=.43\textwidth]{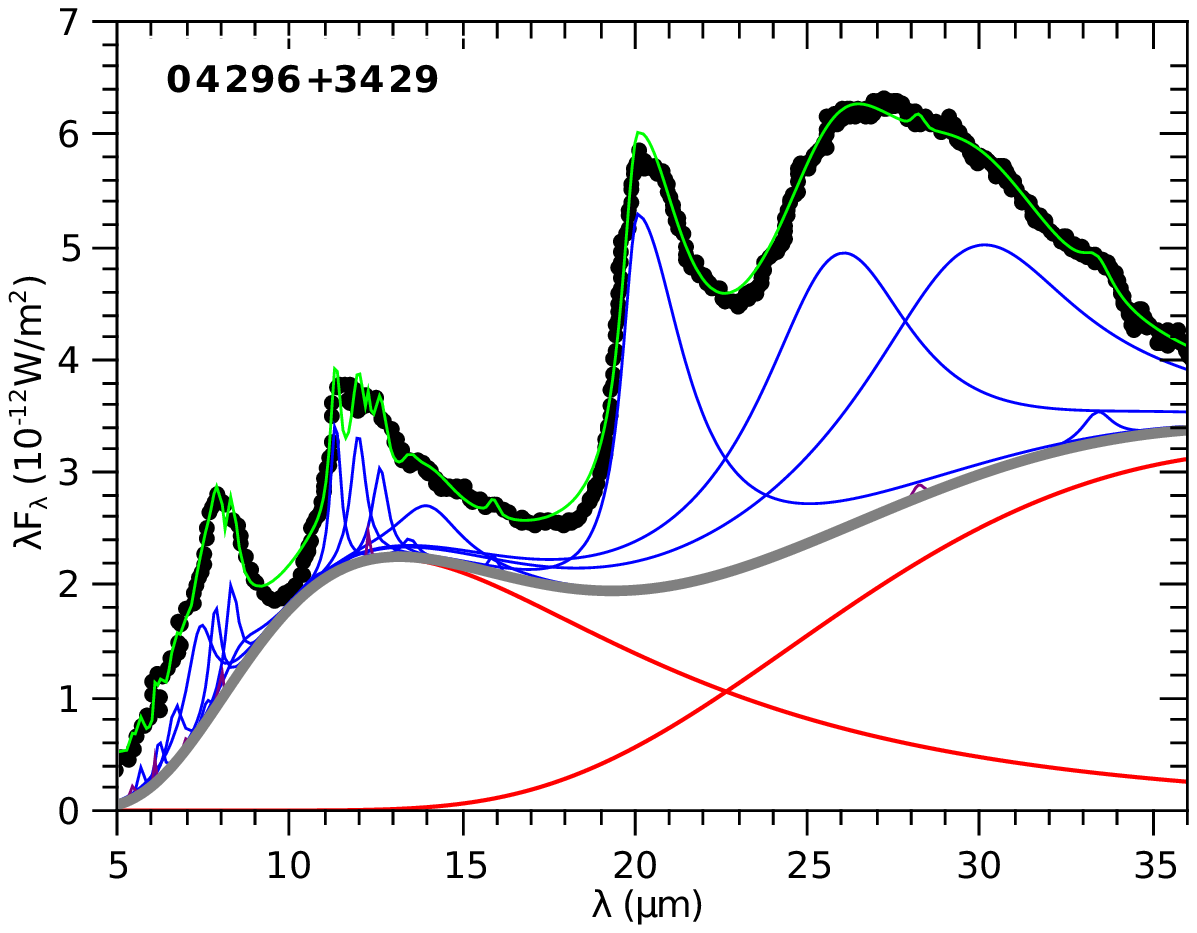} &
    \includegraphics[width=.43\textwidth]{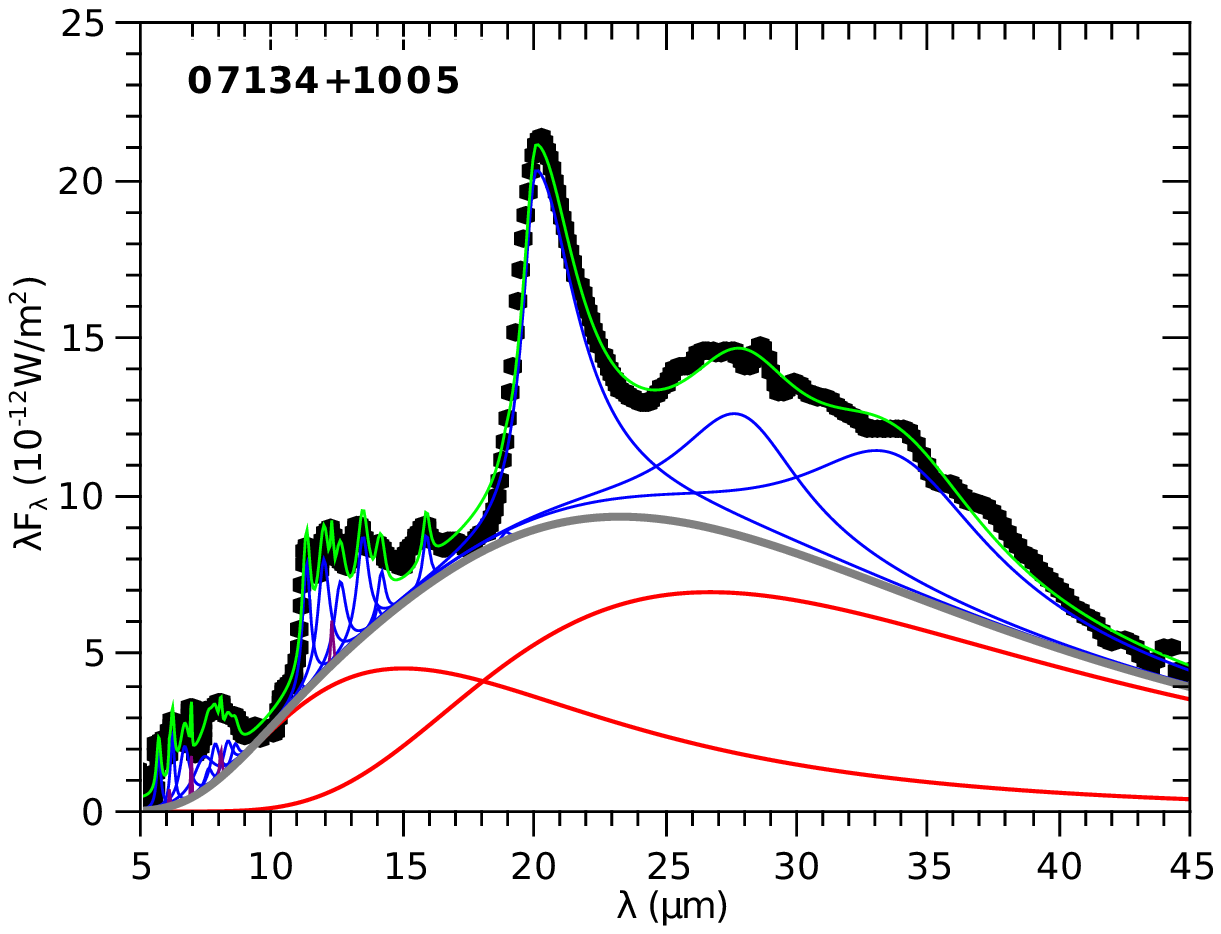} \\
    \includegraphics[width=.43\textwidth]{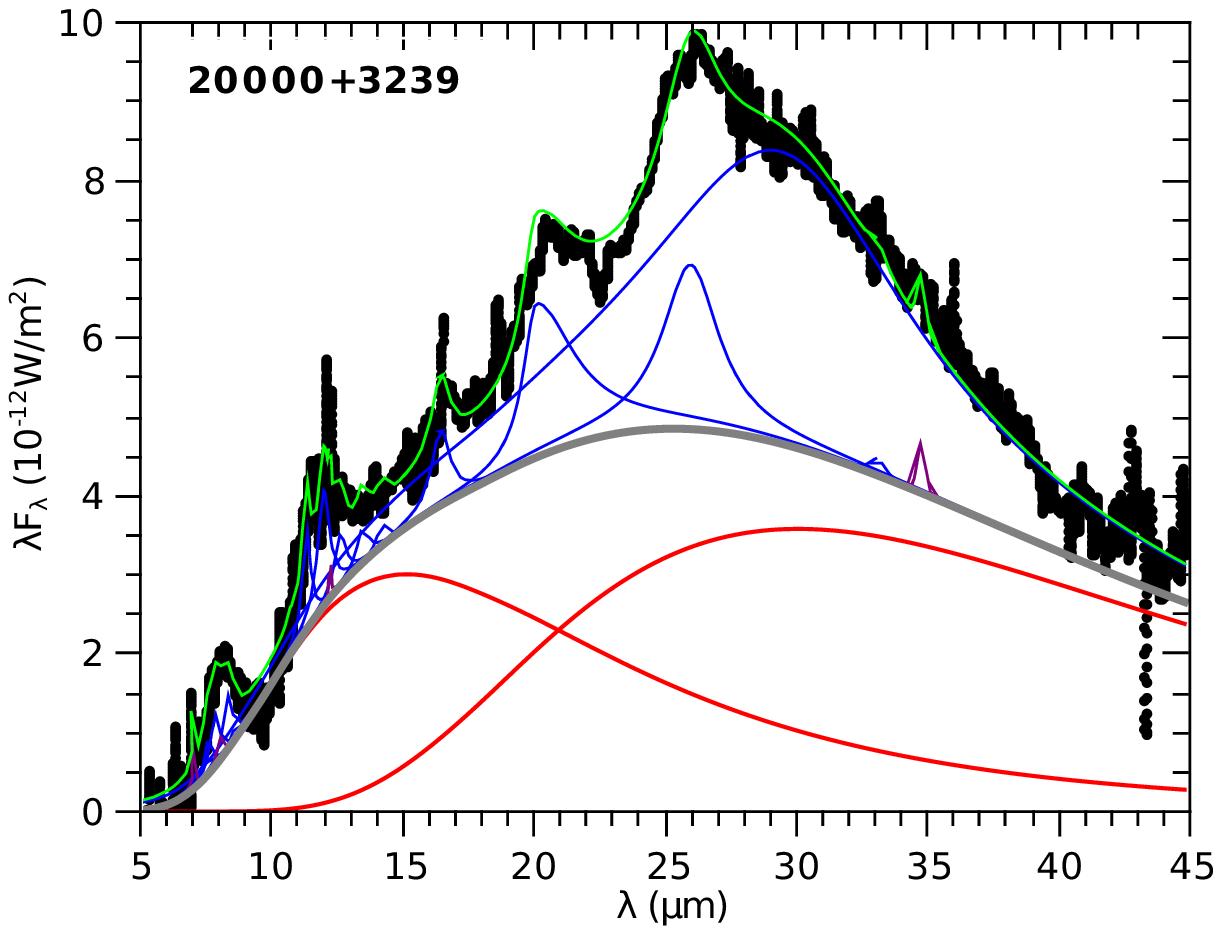} &
    \includegraphics[width=.43\textwidth]{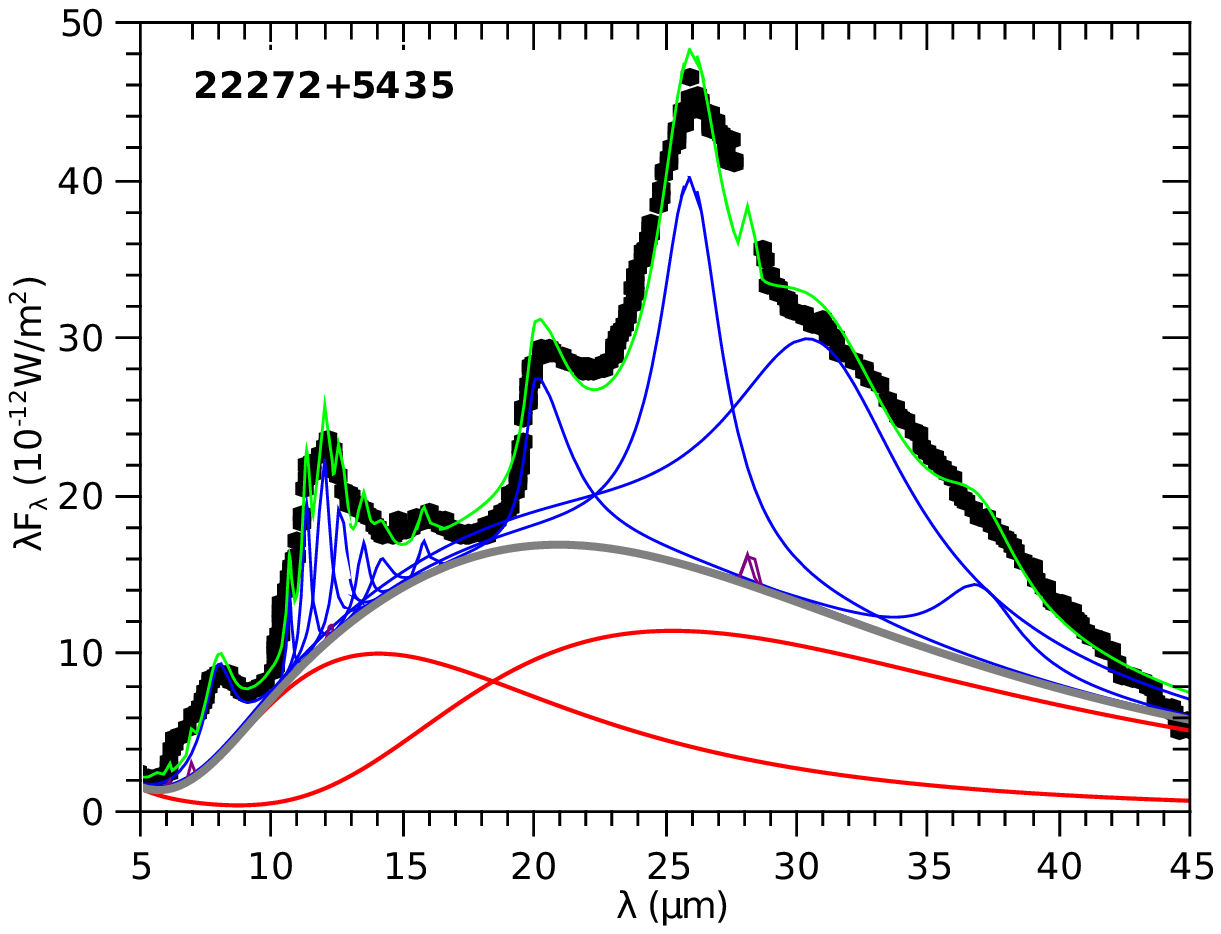} \\ 
  \end{tabular}
\caption{\footnotesize
         \label{fig:drudefit1}
         Decomposition of the observed IR spectra (black points)
         of the 21$\mum$ sources 
         (IRAS\,04296+3429, IRAS\,07134+1005, 
          IRAS\,20000+3239, and IRAS\,22272+5435) 
         into a stellar continuum (pink line), 
         two dust thermal continuum emission components (red lines),
         the 21$\mum$, 30$\mum$ and UIR features (blue lines),
         and the H$_2$ bands (violet lines).
         The decomposition is done with 
         the PAHFIT technique.
         Gray lines are the sum of the stellar 
         and dust thermal continuum.
         Green lines show the resulting model spectra. 
         }
\end{figure*}

\begin{figure*}
\centering
  \begin{tabular}{cc}
    \includegraphics[width=.43\textwidth]{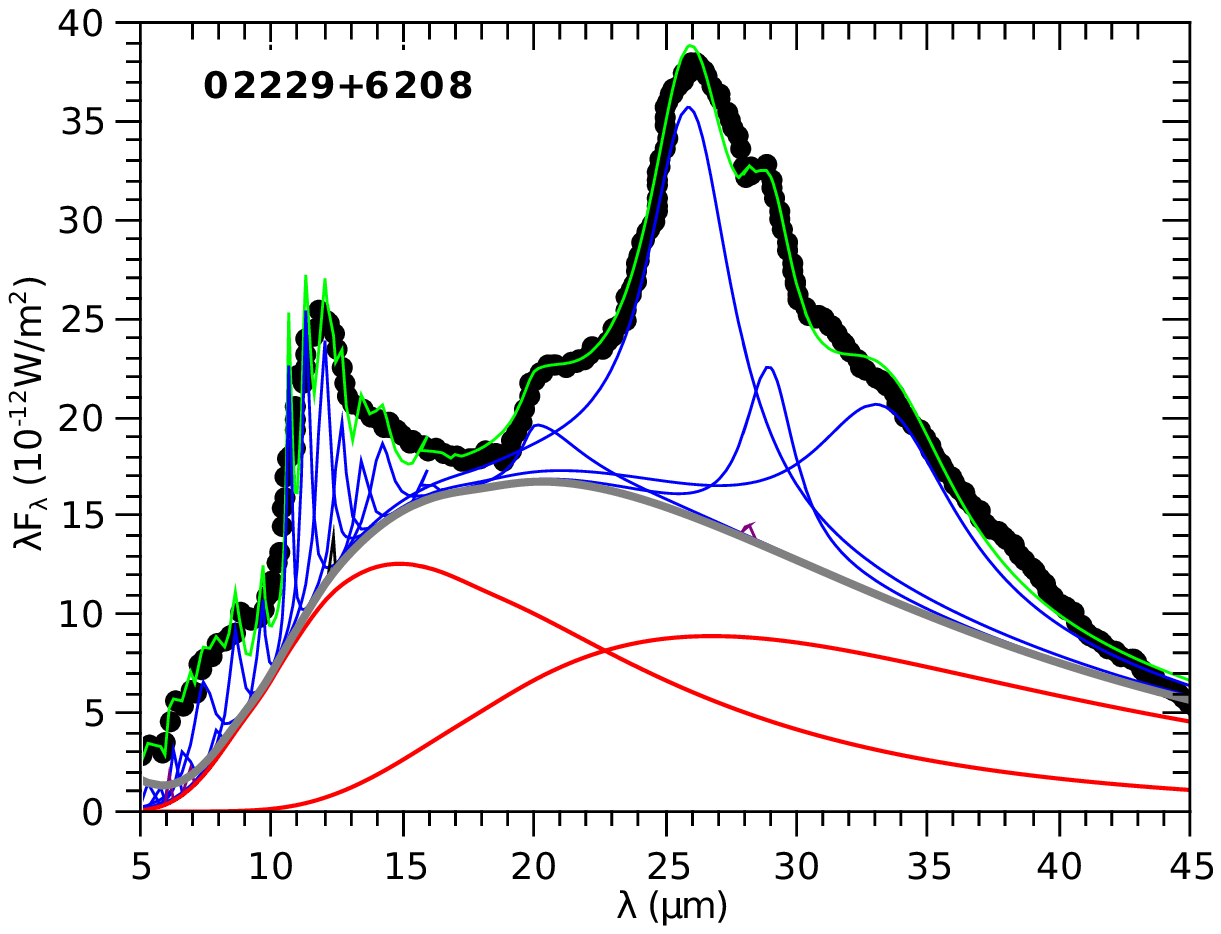} &
    \includegraphics[width=.43\textwidth]{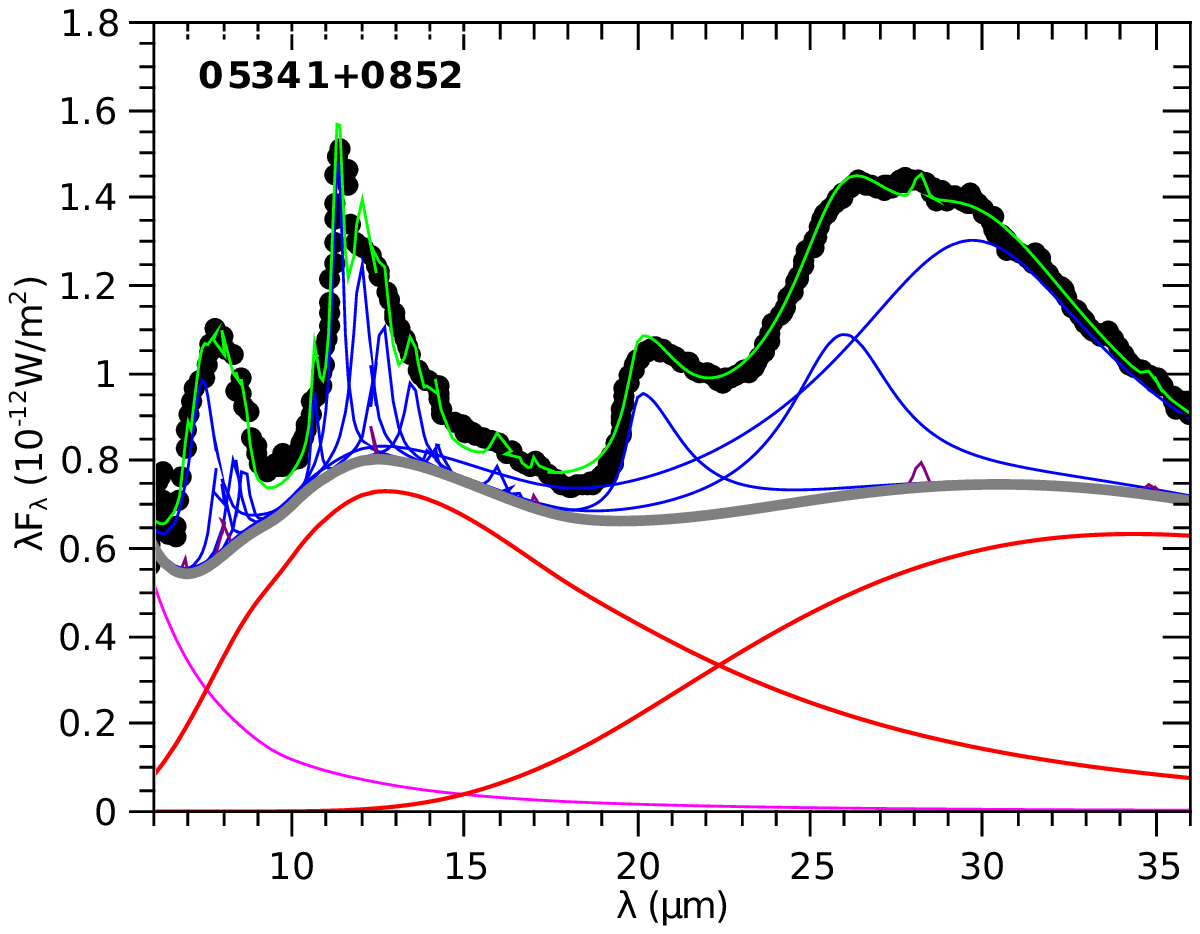} \\
    \includegraphics[width=.43\textwidth]{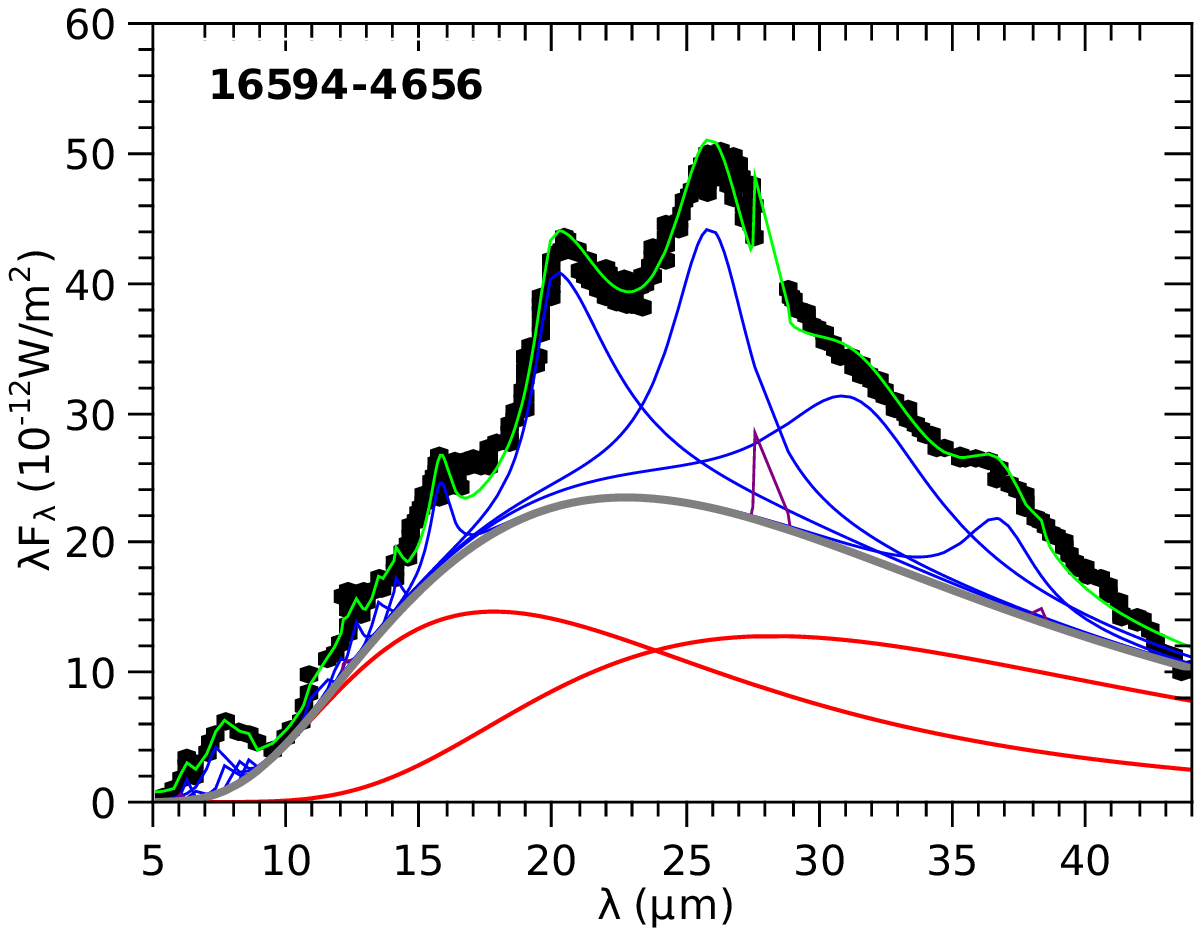} &
    \includegraphics[width=.43\textwidth]{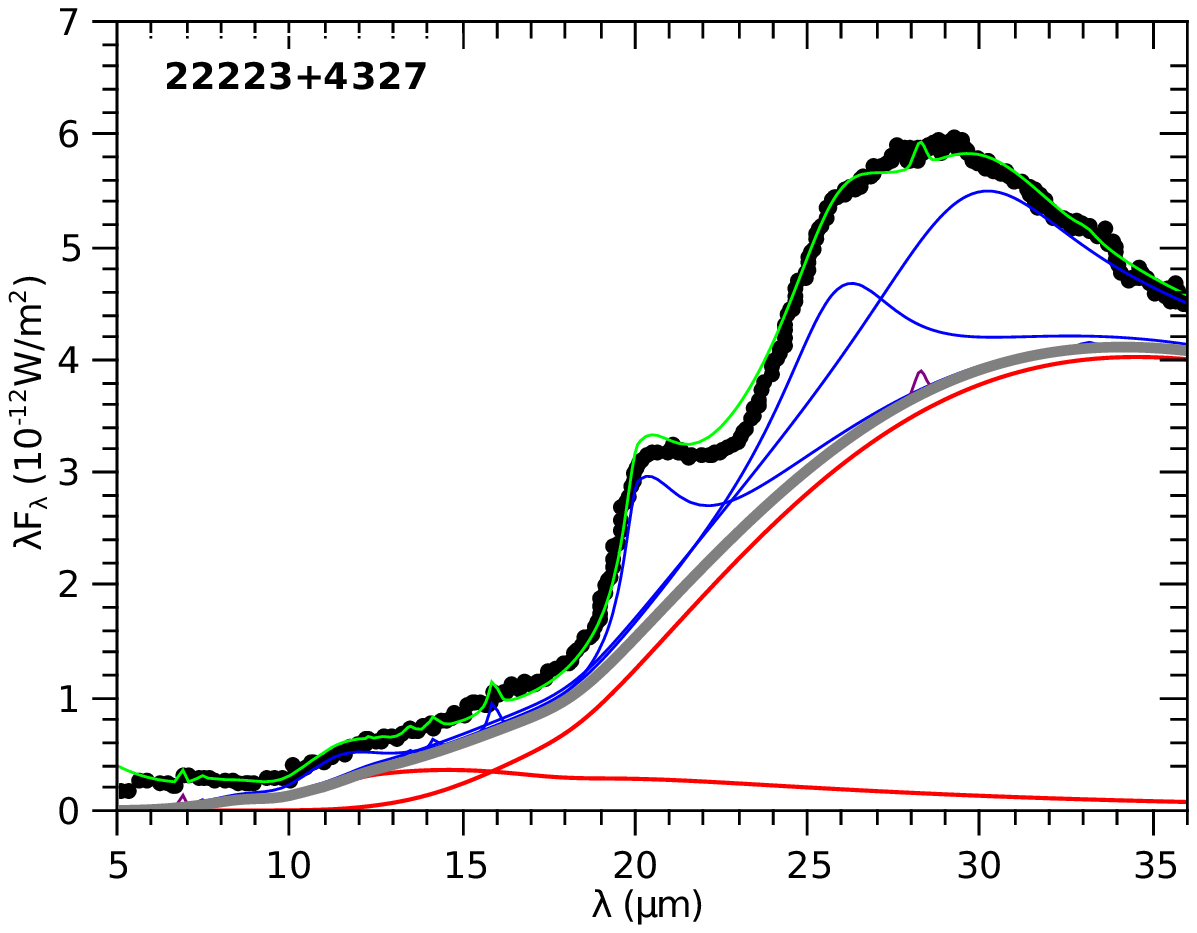}   \\
  \end{tabular}
  \caption{\footnotesize
         \label{fig:drudefit2}
         Same as Figure~\ref{fig:drudefit1}
         but for IRAS\,02229+6208, IRAS\,05341+0852,
         IRAS\,16594-4656, and IRAS\,22223+4327. 
         }
\end{figure*}

\begin{figure*}
\centering
  \begin{tabular}{cc}
    \includegraphics[width=.43\textwidth]{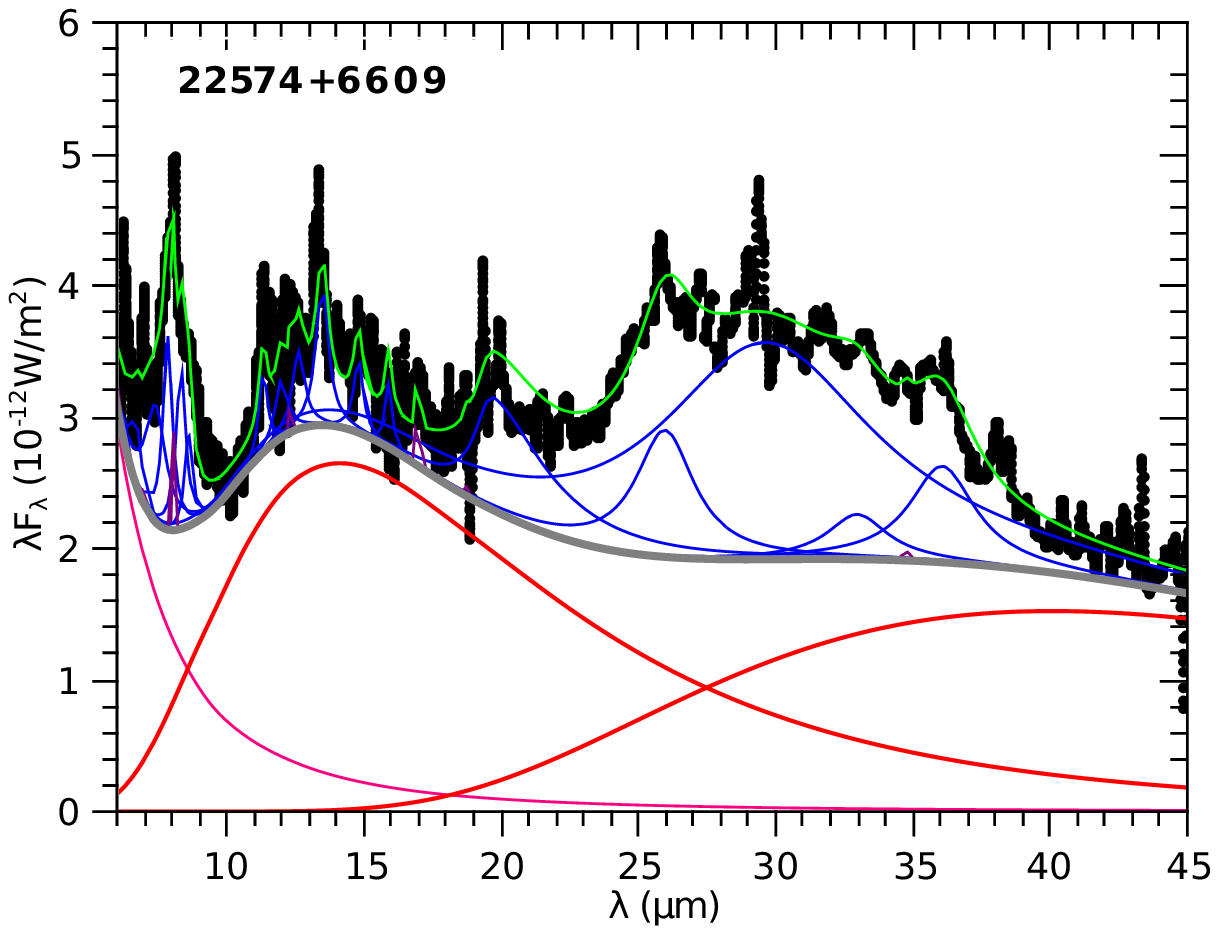} &
    \includegraphics[width=.43\textwidth]{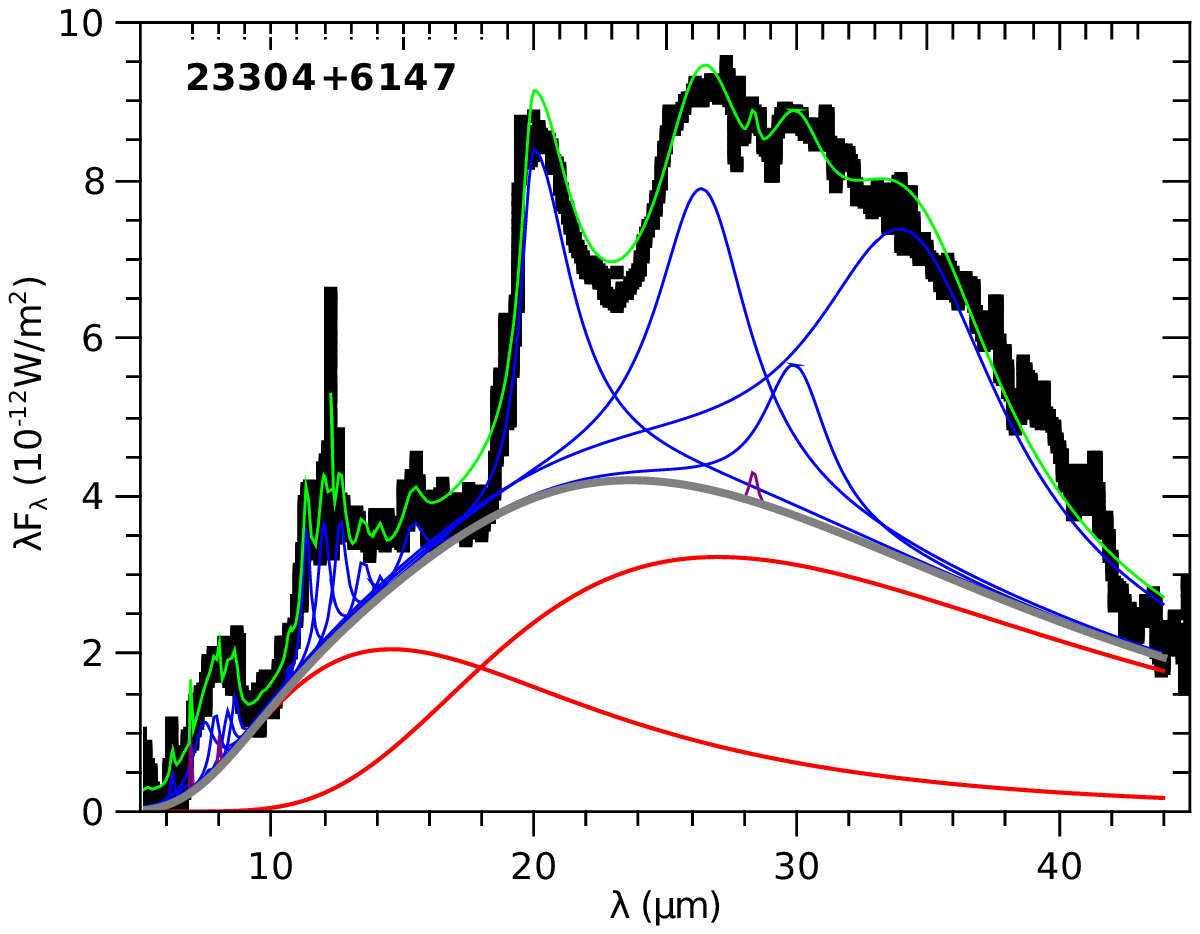}   \\
  \end{tabular}
  \caption{\footnotesize
         \label{fig:drudefit3}
         Same as Figure~\ref{fig:drudefit1}
         but for IRAS\,22574+6609 and IRAS\,23304+6147. 
         }
\end{figure*}

The optimal fitting to the observed spectra is 
achieved through the Levenberg-Marquardt 
least-square algorithm 
implemented in the IDL package.
We are able to obtain reasonably good fits to 
the observed spectra using this technique. 
In Figures~\ref{fig:drudefit1}--\ref{fig:drudefit3}
we show the spectral decomposition fitting 
to the ten sources.
The fluxes emitted by the 21$\mum$, 30$\mum$ 
and UIR features are listed in Table~\ref{tab:drudefit}.
Also listed in Table~\ref{tab:drudefit} is $F_{\rm IR}$,
the dust continuum IR flux 
integrated over the 5--45$\mum$
wavelength range.\footnote{%
   For IRAS\,04296+3429, IRAS\,05341+0852,
   and IRAS\,22223+4327, the IR spectra are 
   not available at $\lambda>35\mum$.
   We derive $F_{\rm IR}$ for these sources by
   extrapolating the observed spectra to 
   $\lambda = 45\mum$.
   }
The uncertainties of the feature strengths 
are estimated using the full covariance matrix 
of the least square parameters as described in
Smith et al.\ (2007).
\begin{table}
\footnotesize{
\caption{\footnotesize
          \label{tab:drudefit} 
          Integrated Fluxes Emitted
          in the UIR ($\Fuir$), 21$\mum$ ($\Ftwenty$), 
          and 30$\mum$ ($\Fthirty$) Features
          as well as the Total Near- to 
          Mid-IR Emission ($\Fir$)
          in the $\sim$\,5--45$\mum$ Wavelength Range, 
          as Derived from the PAHFIT Decomposition Method.
          }
\begin{tabular}{lcccccc}
\hline
\hline
IRAS & Warm Dust & Cold Dust 
        & UIR & $\Ftwenty$ & $\Fthirty$ 
        & $F_{\rm IR}$ \\
Sources & $T_W$ (K) & $T_C$ (K)
             &($10^{-12}\W\m^{-2}$) 
             & ($10^{-12}\W\m^{-2}$)
             & ($10^{-12}\W\m^{-2}$) 
             & ($10^{-12}\W\m^{-2}$)\\
\hline
02229+6208 & 160 & 90 & 5.78$\pm$0.016 & 0.322$\pm$0.002 & 9.10$\pm$0.138 & 25.6$\pm$1.95\\
04296+3429 & 185 & 61 & 0.870$\pm$0.006 & 0.302$\pm$0.002 & 1.62$\pm$0.012 & 7.15$\pm$0.23 \\
05341+0852 & 190 & 70 & 0.283$\pm$0.005 & 0.023$\pm$0.008 & 0.35$\pm$0.002 & 1.66$\pm$0.07 \\
07134+1005 & 160 & 90 & 1.84$\pm$0.017 & 1.40$\pm$0.002 & 2.88$\pm$0.015 & 17.3$\pm$1.60 \\
16594-4656 & 135 & 78 & 3.25$\pm$0.014 & 2.57$\pm$0.002 & 9.97$\pm$0.080 & 43.6$\pm$2.90 \\
20000+3239 & 158 & 80 & 0.522$\pm$0.009 & 0.185$\pm$0.001 & 2.63$\pm$0.019 & 7.29$\pm$1.50 \\
22223+4327 & 150 & 70 & 0.158$\pm$0.005 & 0.114$\pm$0.001 & 0.986$\pm$0.014 & 5.42$\pm$0.30 \\
22272+5435 & 165 & 90 & 4.22$\pm$0.036 & 0.840$\pm$0.010 & 11.4$\pm$0.135 & 35.8$\pm$2.20 \\
22574+6609 & 170 & 60 & 0.734$\pm$0.008 & 0.144$\pm$0.002 & 1.20$\pm$0.006 & 4.61$\pm$0.69 \\
23304+6147 & 165 & 89 & 0.767$\pm$0.011 & 0.416$\pm$0.001 & 2.88$\pm$0.010 & 8.90$\pm$0.83 \\

\hline
\end{tabular}
\tablecomments{
    For IRAS\,04296+3429, IRAS\,05341+0852,
    and IRAS\,22223+4327, the IR spectra are 
    not available at $\lambda>35\mum$.
    We derive $F_{\rm IR}$ for these sources by
    extrapolating the observed spectra to 
   $\lambda = 45\mum$.
     %
     }
}
\end{table}

\section{Interrelations among the 21$\mum$, 30$\mum$ 
         and UIR Features \label{sec:2130uir}}
%


Let $\Mtwenty$, $\Mthirty$, and $\Muir$ respectively be 
  the mass of the carrier of the 21$\mum$, 30$\mum$, and UIR features.
  Let $\kappatwenty$, $\kappathirty$, and $\kappauir$ respectively be 
  the ultraviolet/visible mass absorption coefficient 
  of the carrier of the 21$\mum$, 30$\mum$, and UIR features.  
  It is reasonable to assume that 
  the total masses of the carriers 
  of these unidentified features
  ($\Mtwenty$, $\Mthirty$, and $\Muir$)  
  are proportional to the total mass of 
  the bulk dust ($\Mdust$) in the shell,
  although the proportional factor may 
  differ from one to another.   
  Let $\fractwenty$, $\fracthirty$, and $\fracuir$ respectively be 
  the ratio of the mass of the carrier of the 21$\mum$, 30$\mum$, 
  and UIR features to the mass of the bulk dust in a 21$\mum$ source.
  With $\Fir\propto \left(\Mdust\times\Lstar\right)$ we have 
  $\Ftwenty \propto \left(\Mtwenty\times\kappatwenty\times\Lstar\right)
  \propto \left(\fractwenty\times\Mdust\times\Lstar\right)
  \propto \left(\fractwenty\times\Fir\right)$.
  Similarly, we have $\Fthirty \propto \left(\fracthirty\times\Fir\right)$
  and $\Fuir \propto \left(\fracuir\times\Fir\right)$.
  Therefore, even if $\fracthirty$ does not correlate with 
  $\fractwenty$ or $\fracuir$, $\Fthirty$ could still correlate
  with $\Ftwenty$ and $\Fuir$ through $\Fir$.
  Therefore, for a physically reasonable correlation study 
  of the 21$\mum$, 30$\mum$ and UIR features, one need to
  normalize $\Ftwenty$, $\Fthirty$ and $\Fuir$ by $\Fir$,
  in order to cancel out their common proportionality to $\Fir$.

We explore the interrelations among the 21$\mum$, 30$\mum$
and UIR features. We normalize the fluxes emitted from these
features ($\Ftwenty$, $\Fthirty$, $\Fuir$)
by $\Fir$, the total near- to mid-IR emission
in the wavelength range of 5--45$\mum$.
This is to see what fractions of 
the total near- to mid-IR luminosity 
are being carried by these features. 
\begin{figure*}
\centering
\vspace{-0.5cm}
\includegraphics[width=8.0cm]{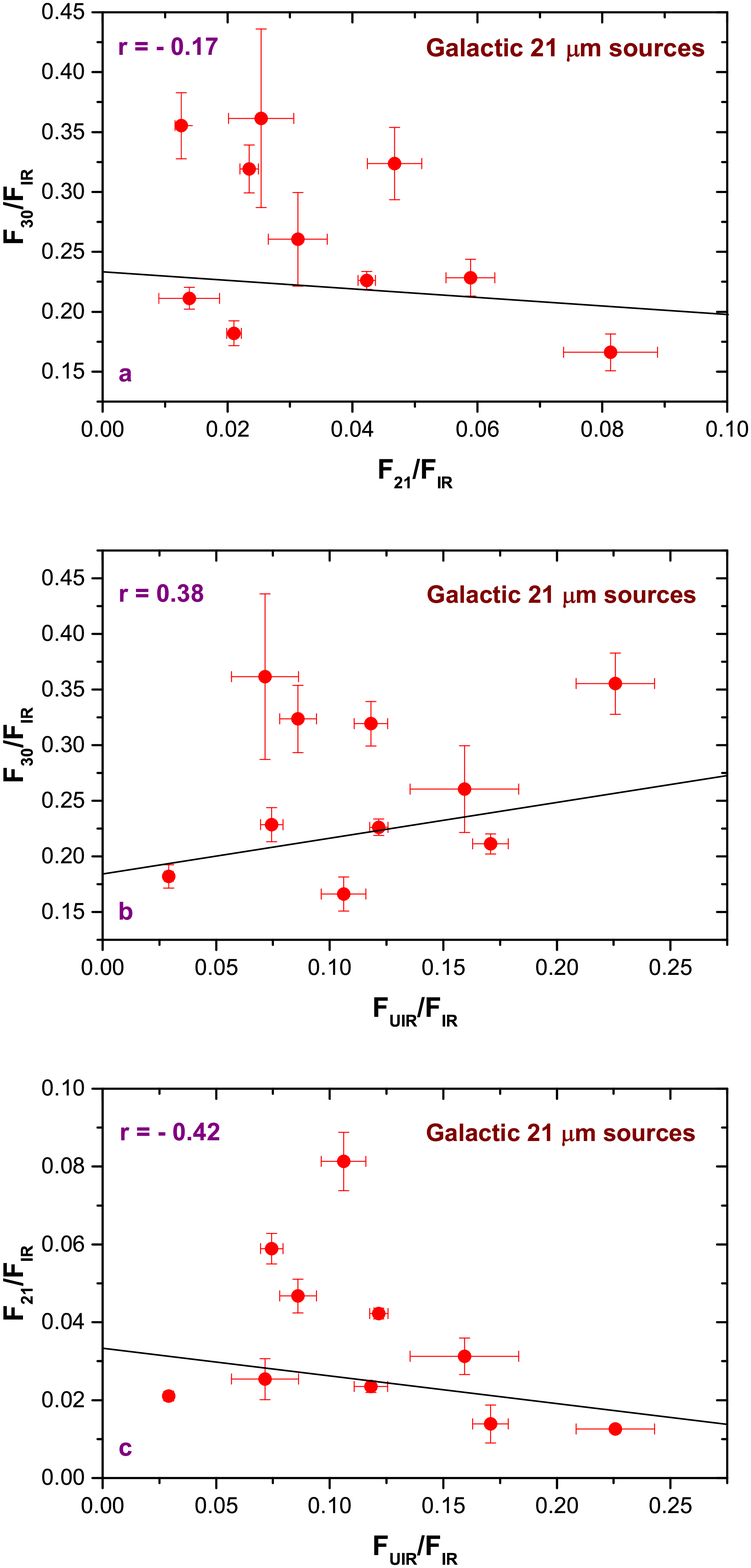}
\vspace{-0.5cm}
\caption{\footnotesize
         \label{fig:2130uir_pahfit}
         Interrelations among the 21$\mum$, 30$\mum$,
         and UIR features of ten Galactic 21$\mum$ sources:
         (a) $\Fthirty$ vs. $\Ftwenty$,
         (b) $\Fthirty$ vs. $\Fuir$, and 
         (c) $\Ftwenty$ vs. $\Fuir$.
         All quantities are normalized by $\Fir$,
         the total near- to mid-IR emission obtained
         by {\it ISO} or {\it Spitzer} 
         in the 5--45$\mum$ wavelength range,
         to cancel out their common proportionality 
         to $\Fir$. 
         The fluxes emitted in the 21$\mum$, 30$\mum$,
         and UIR features are derived from the PAHFIT
         decomposition method.
         }
\end{figure*}


We show in Figure~\ref{fig:2130uir_pahfit}
the interrelations of the fractional fluxes
of the 21$\mum$, 30$\mum$ and UIR features.
It is clear from Figure~\ref{fig:2130uir_pahfit} 
that these features are not correlated. 
This differs from Cerrigone et al.\ (2011) 
who reported a positive correlation between 
$\Fuir$ and $\Ftwenty$ and suggested that 
the abundances of the carrier of the UIR 
bands and of the carrier of the 21$\mum$ 
feature may be linked together, 
possibly because of a common formation process.
We attribute the difference to 
the fact that Cerrigone et al.\ (2011)
did not normalize $\Fuir$ and $\Ftwenty$ by $\Fir$.
We argue that the correlation between 
$\Ftwenty$ and $\Fuir$ found by 
Cerrigone et al.\ (2011) may 
merely reflect the fact that both 
$\Ftwenty$ and $\Fuir$ correlate with $\Fir$
(i.e., with $\Mtwenty\propto\Mdust$, $\Mthirty\propto\Mdust$,
and $\Muir\propto\Mdust$, even if $\fracthirty$ does not correlate 
with $\fractwenty$ or $\fracuir$, $\Mthirty$ could still correlate
with $\Mtwenty$ and $\Muir$). 
\section{Implications\label{sec:implications}}
\vspace{-1mm}
The UIR bands are commonly attributed 
to PAH molecules. 
If the 21$\mum$ feature is due to 
large PAH clusters (Buss et al.\ 1990), 
$\Ftwenty$ and $\Fuir$ should be somewhat correlated: 
the fast stellar wind from the central star 
during the PPN phase might shatter PAH clusters 
into PAH molecules, possibly obliterating 
the carrier of the 21$\mum$ feature.
The lack of correlation between $\Fuir$ 
and $\Ftwenty$ indicates that the 21$\mum$ 
feature is probably not due to 
large PAH clusters. 

  
Papoular (2000, 2011) hypothesized that macromolecules 
with thiourea functional groups 
and aliphatic chains (made of CH$_2$ groups, 
oxygen bridges and OH groups)
attached to various carbonaceous structures 
(mainly compact and linear aromatic clusters)
are responsible for the 21$\mum$, 30$\mum$, and UIR features,
with thiourea emitting the 21$\mum$ feature, 
the aliphatic chains producing the 30$\mum$ feature,
and the aromatic clusters giving rise to the UIR bands.
If this is the case, one would expect 
the 21$\mum$, 30$\mum$ and UIR features 
to be positively correlated. 
The lack of any positive correlation 
among these features as demonstrated in \S2 
does not support the hypothesis of 
Papoular (2000, 2011).

%
%
Volk et al.\ (2011) reported an anti-correlation 
between the 30$\mum$ and UIR features 
for the Magellanic Cloud 21 $\mum$ sources.
They postulated there may be some competition in 
forming the UIR and 30$\mum$ feature carriers.
Indeed, if the 30$\mum$ feature is due to 
hydrogenated amorphous carbon (HAC), 
such an anti-correlation could result from 
radiation-induced decomposition of HAC grains 
into PAHs (Scott et al.\ 1997) and/or stellar 
wind-induced shattering of HAC grains into PAHs.
However, as shown in Figure~\ref{fig:2130uir_pahfit}
for the Galactic 21$\mum$ sources
the 30$\mum$ and UIR features 
do not seem to be anti-correlated.
%
%
%
%
%
%

   Spectroscopically, the Magellanic Cloud 21$\mum$ sources
   differ appreciably from the Galactic 21$\mum$ sources
   (see Volk et al.\ 2011):
   (1) the average 21$\mum$ feature strength 
       for the Magellanic Cloud objects 
       is smaller than that observed in the Galaxy;
   (2) the 30$\mum$ feature of the Magellanic Cloud sources  
       carries less of the total mid-IR emission 
       than is normally the case for the Galactic 21$\mum$ sources;
   (3) the UIR features of the Magellanic Cloud 21$\mum$ sources
       are typically twice as strong as in the Galactic sources,
       with the peak of the UIR emission often being higher than 
       that of the 30$\mum$ feature; and
     (4) 
         the Magellanic Cloud objects have more structure 
         in their UIR spectra and show more variation 
         in the shapes and strengths of the 7.7 and 8.6$\mum$ features 
         than is observed in the Galactic 21$\mum$ sources;
         more particularly, most of the Galactic 21$\mum$ sources have 
         ``unusual'' UIR properties 
         compared to the typical interstellar UIR bands
         (e.g., the Galactic 21$\mum$ sources 
         have a broad 8$\mum$ feature 
         rather than separately resolved 
         7.7/8.6$\mum$ UIR features) 
         while the Magellanic Cloud 21$\mum$ sources 
         show more ``normal'' looking UIR features 
         (see Peeters et al.\ 2002).
         %
%
The remarkable difference in the spectral 
appearance of the UIR features 
between the Galactic 
and Magellanic Cloud 21$\mum$ sources
would potentially hold important clues
for understanding how the physical 
and chemical conditions 
affect the chemical structures of 
the carriers of these unidentified features.

An unique characteristic of the mysterious 21$\mum$ feature
is that it is almost {\it exclusively} seen in the short-lived 
PPN phase of stellar evolution and always accompanied by 
the equally mysterious 30$\mum$ feature and UIR bands, 
while the 30$\mum$ feature is commonly observed
in all late stages of stellar evolution 
from the AGB through the PPNe to the PNe phase
(the UIR features are seen in PPNe and PNe, 
but not in AGB stars). 
This implies that the carriers of the 21$\mum$ feature
are likely formed during the so-called superwind phase
(e.g., see von Helden et al.\ 2000), 
which is a phase of high mass-loss where AGB stars 
lose the remaining stellar envelope, terminating their life 
on the AGB (see Renzini 1981).
They may also be easily destroyed by the highly energetic 
photons available in PNe.
We also note that while the UIR features 
are commonly seen in PPNe, PNe, and the ISM, 
they are usually not seen in the mid-IR spectra 
of carbon-rich AGB stars.\footnote{%
  The few carbon stars that display 
  the UIR features all have a hot companion 
  that emits UV photons 
  (Speck \& Barlow 1997, Boersma et al.\ 2006).
   It has been argued that the carriers of
   the UIR features (e.g., PAHs) are present 
   in carbon stars but are simply not excited 
   sufficiently to emit at mid-IR due to lack 
   of UV photons. 
   However, Li \& Draine (2002) have shown that
   the excitation of PAHs does not require UV photons,
   visible and near-IR photons are capable of exciting PAHs
   to emit at the UIR bands.
   The visible/near-IR cross sections measured
   by Mattioda et al.\ (2005) for PAH ions imply that PAHs 
   can be excited by the soft stellar photons from C stars. 
   Therefore, the absence of the UIR bands in C stars
   places a limit on the abundance of small PAHs 
   in these outflows.
   }

Finally, we also note that, in addition
to the variable 30$\mum$ feature (in contrast to
the invariable 21$\mum$ feature),\footnote{%
   Hony et al.\ (2002) found that the peak wavelength 
   of the 30$\mum$ feature shifts 
   from $\simali$26$\mum$ in some AGB stars 
   to $\simali$38$\mum$ in PNs.
   }
the overall shape of the {\it Spitzer} spectra of 
the 21$\mum$ sources show also a lot of diversity,
from having almost no flux at $\simali$10$\mum$ 
(like IRAS 22223-4327) to having a very flat spectrum 
(like IRAS 05341+0852, IRAS 22574+6609). 
We do not know what causes this diversity.
We have examined the stellar luminosity, 
and the optical depth,
the expansion history,
and the inner and outer edges of the dust shell.
They do not seem to be responsible for this overall
spectral shape diversity.

\begin{table}
\footnotesize{
\caption{\footnotesize
         \label{tab:SplineFlux}
         Same as Table~\ref{tab:drudefit}
         but with the Dust Continuum 
         Fitted with a Spline Function
         (See \S\ref{sec:continuum}, and
          Figures~\ref{fig:spline_fit1},\ref{fig:spline_fit2}).
          }
\begin{center}
\begin{tabular}{lcccc}
\hline
\hline
IRAS    & UIR & $\Ftwenty$ & $\Fthirty$ 
        & $F_{\rm IR}$ \\
Sources &($10^{-12}\W\m^{-2}$) 
        & ($10^{-12}\W\m^{-2}$)
        & ($10^{-12}\W\m^{-2}$) 
        & ($10^{-12}\W\m^{-2}$)\\
\hline
02229+6208  & 5.05$\pm$0.57  & 0.34$\pm$0.02   & 7.00$\pm$0.58  &  25.6$\pm$1.95 \\
04296+3429  & 0.87$\pm$0.06  & 0.35$\pm$0.02   & 0.63$\pm$0.06  &  7.15$\pm$0.23 \\
05341+0852  & 0.28$\pm$0.04  & 0.029$\pm$0.002 & 0.15$\pm$0.01  &  1.66$\pm$0.07 \\
07134+1005  & 1.90$\pm$0.30  & 1.62$\pm$0.06   & 2.07$\pm$0.25  &  17.3$\pm$1.60 \\
16594-4656  & 3.46$\pm$0.41  & 2.02$\pm$0.041  & 4.10$\pm$0.07  &  43.6$\pm$2.90 \\
20000+3239  & 0.55$\pm$0.09  & 0.19$\pm$0.008  & 1.56$\pm$0.11  &  7.29$\pm$1.50 \\
22223+4327  & 0.15$\pm$0.02  & 0.10$\pm$0.004  & 0.57$\pm$0.05  &  5.42$\pm$0.30 \\
22273+5435  & 4.12$\pm$0.61  & 0.89$\pm$0.05   & 11.0$\pm$0.71  &  35.8$\pm$2.20 \\
22574+6609  & 0.72$\pm$0.10  & 0.11$\pm$0.005  & 0.75$\pm$0.12  &  4.61$\pm$0.69 \\
23304+6147  & 0.76$\pm$0.09  & 0.54$\pm$0.045  & 2.70$\pm$0.36  &  8.90$\pm$0.83 \\
\hline
\end{tabular}
\end{center}
}
\end{table}

\begin{figure*}
\centering
  \begin{tabular}{cc}
    \includegraphics[width=.43\textwidth]{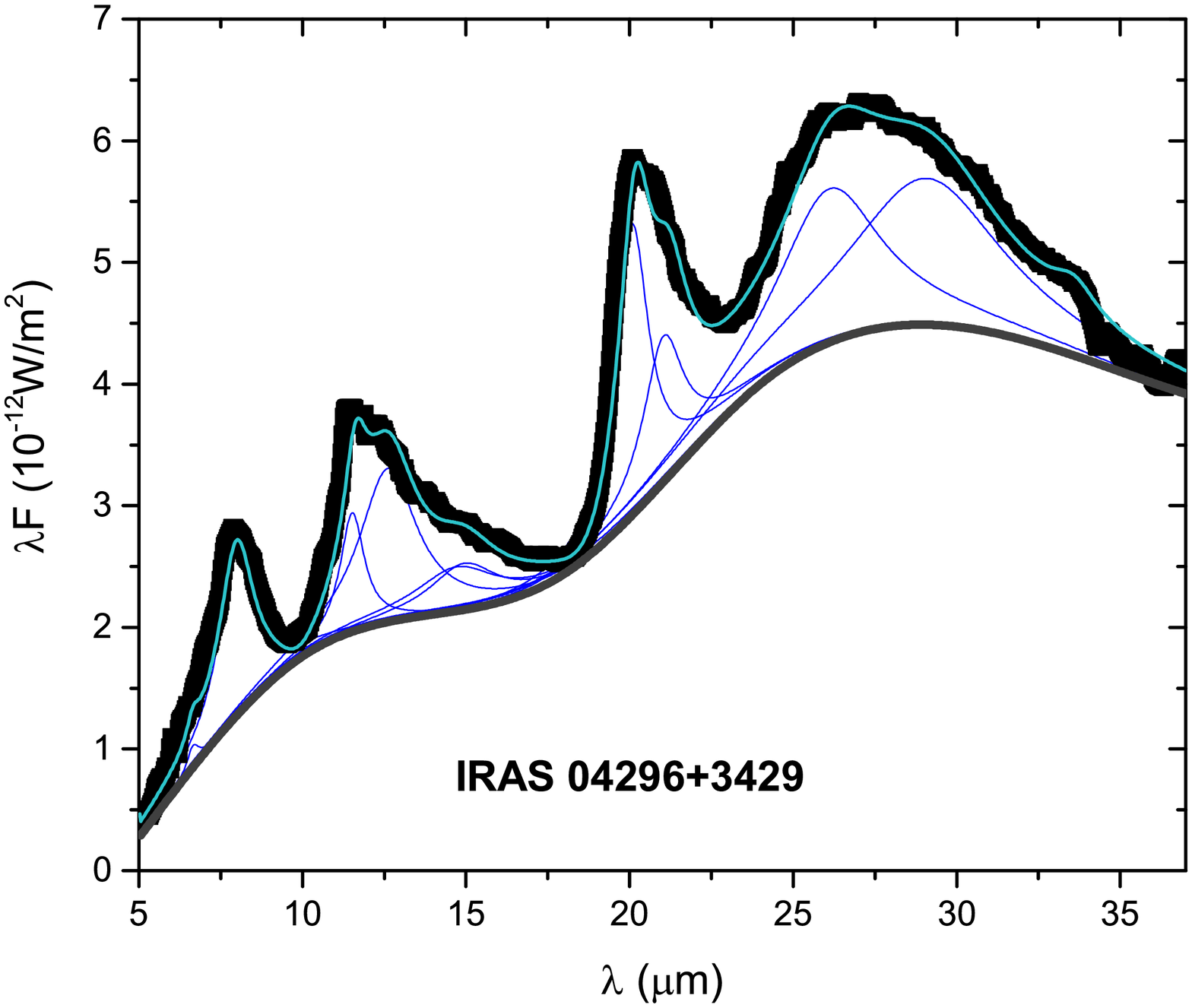} &
    \includegraphics[width=.43\textwidth]{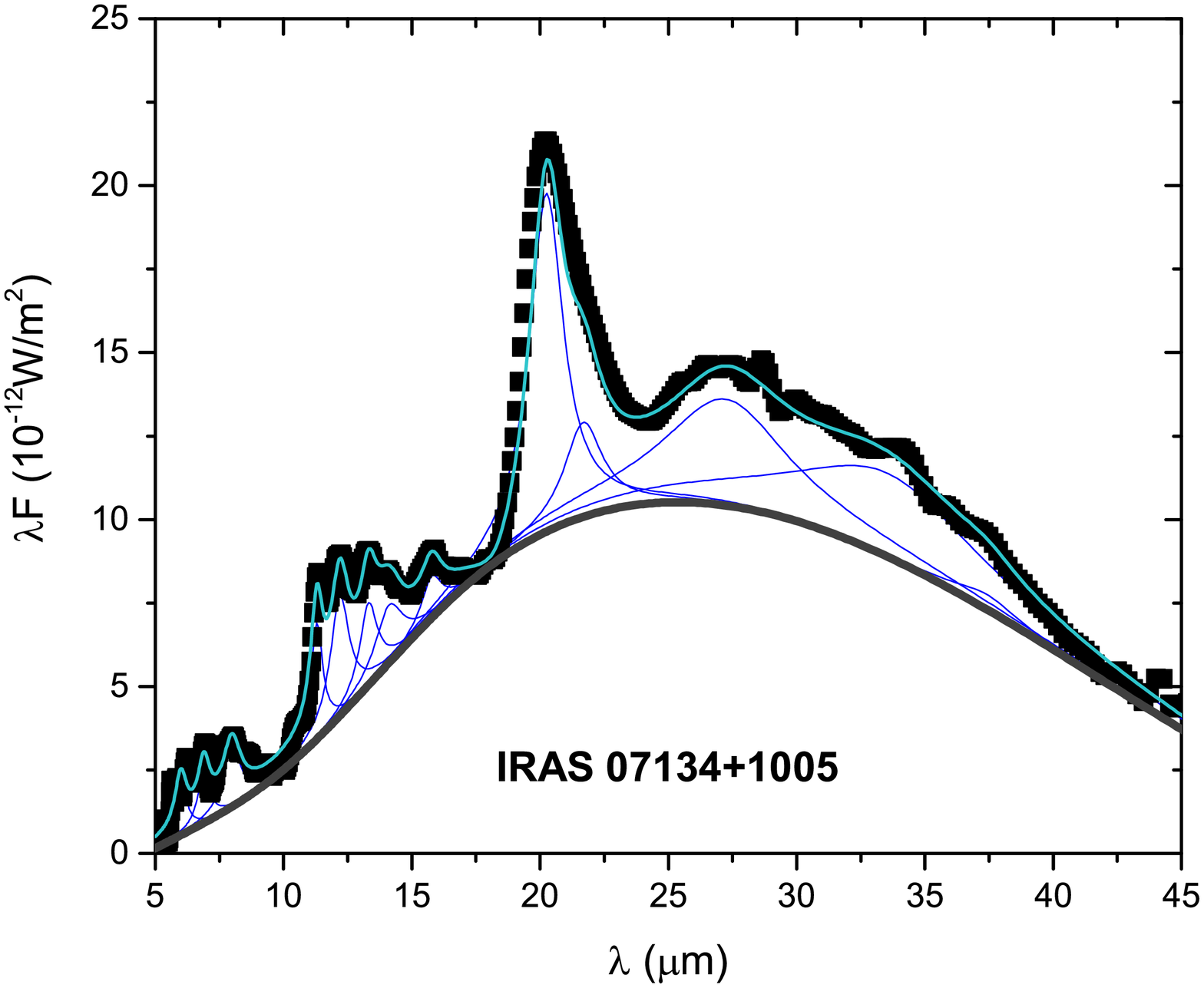} \\
    \includegraphics[width=.43\textwidth]{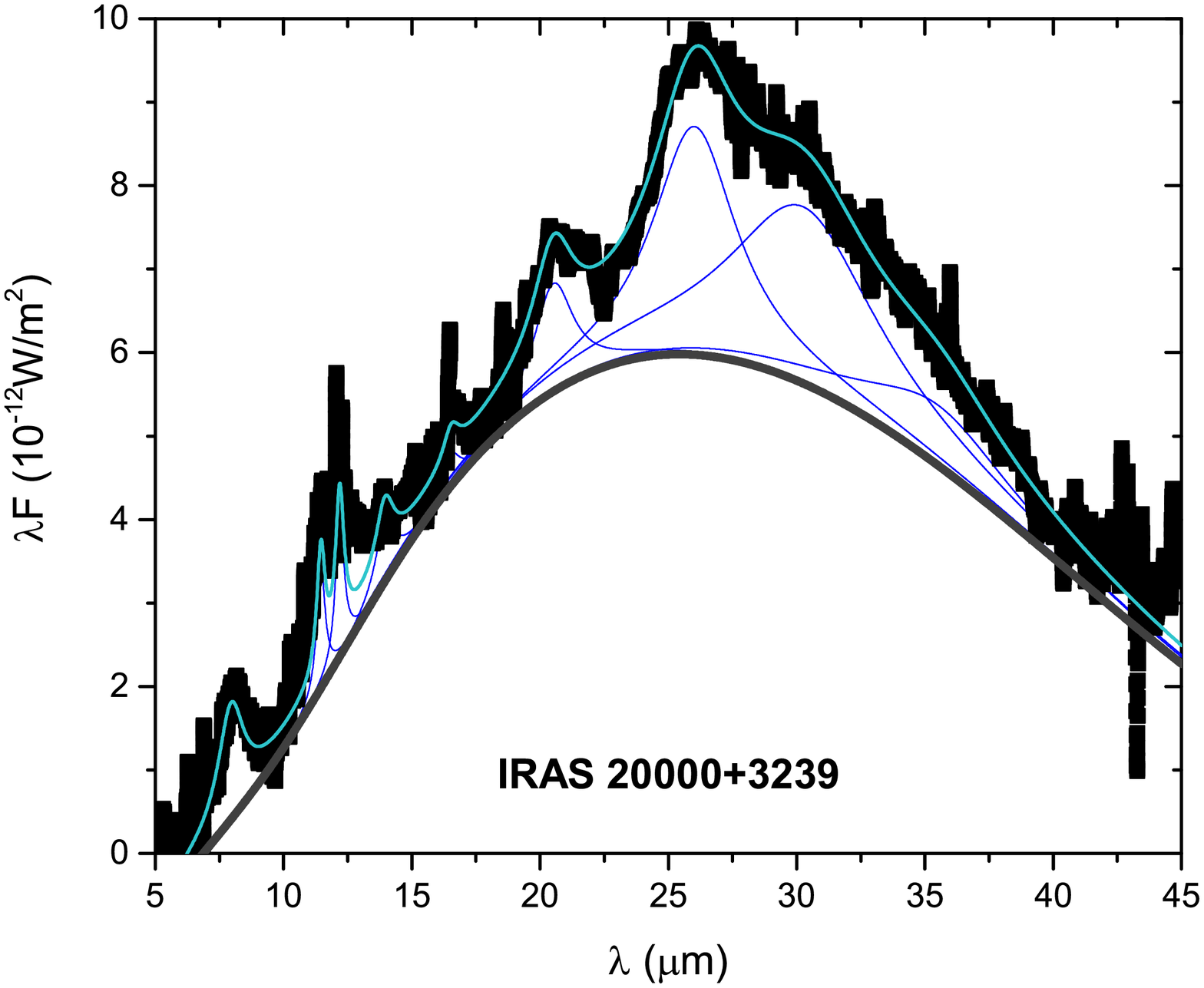} &
    \includegraphics[width=.43\textwidth]{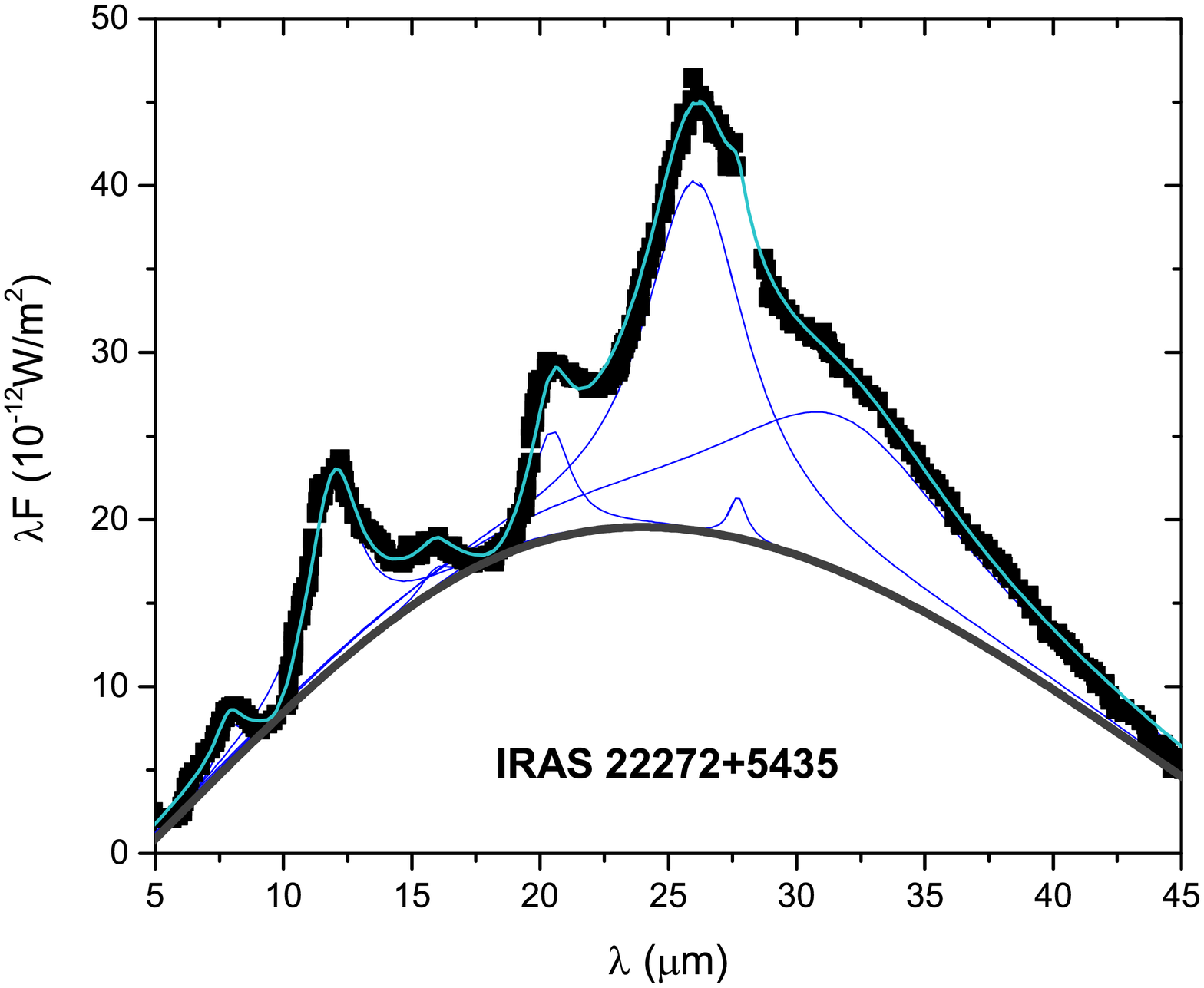}   \\
  \end{tabular}
\caption{\footnotesize
         \label{fig:spline_fit1}
         Same as Figure~\ref{fig:drudefit1}   
         but with the dust continuum 
         fitted with a spline function
         instead of two modified blackbodies. 
         }
\end{figure*}

\begin{figure*}
\centering
  \begin{tabular}{cc}
    \includegraphics[width=.43\textwidth]{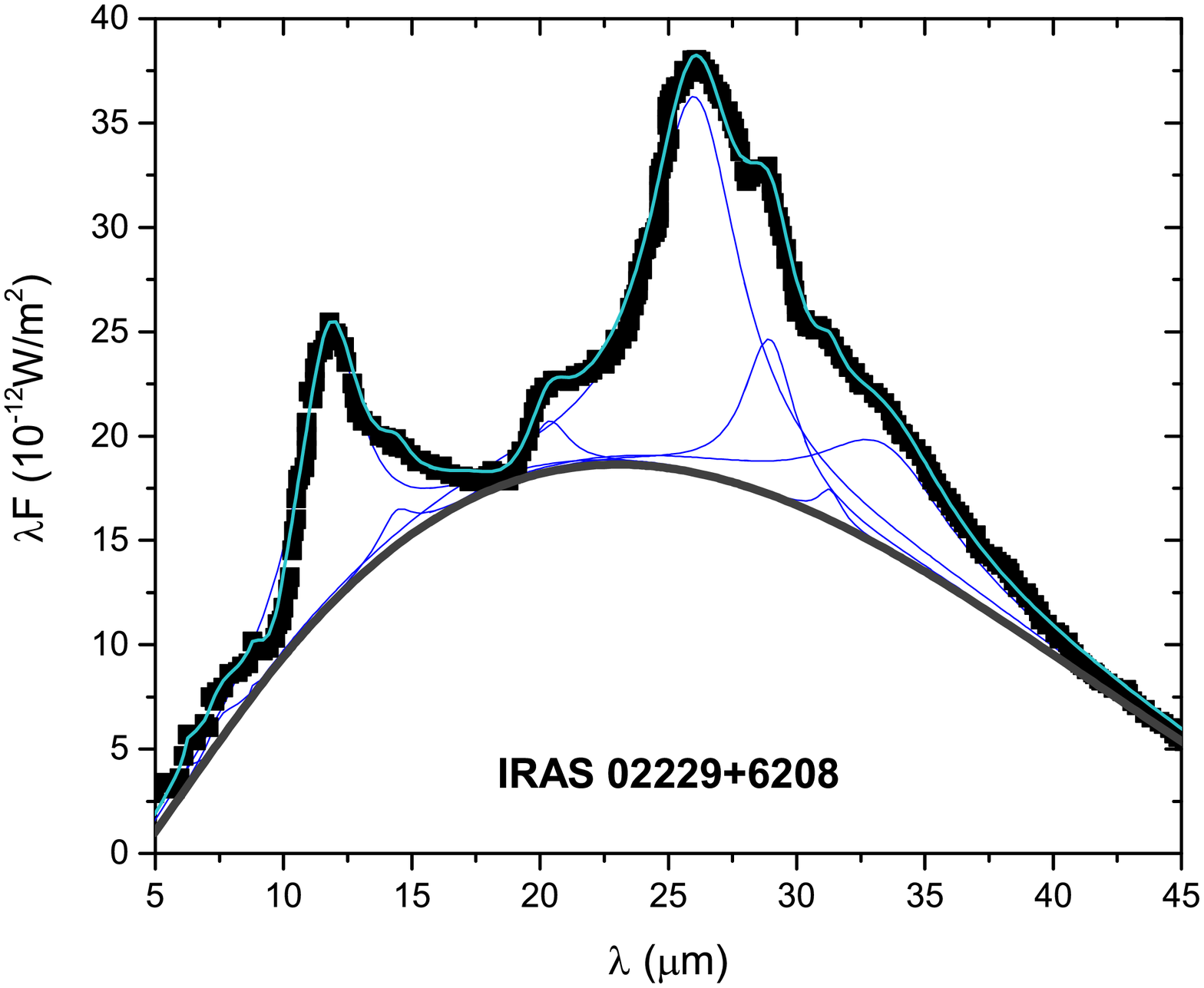} &
    \includegraphics[width=.43\textwidth]{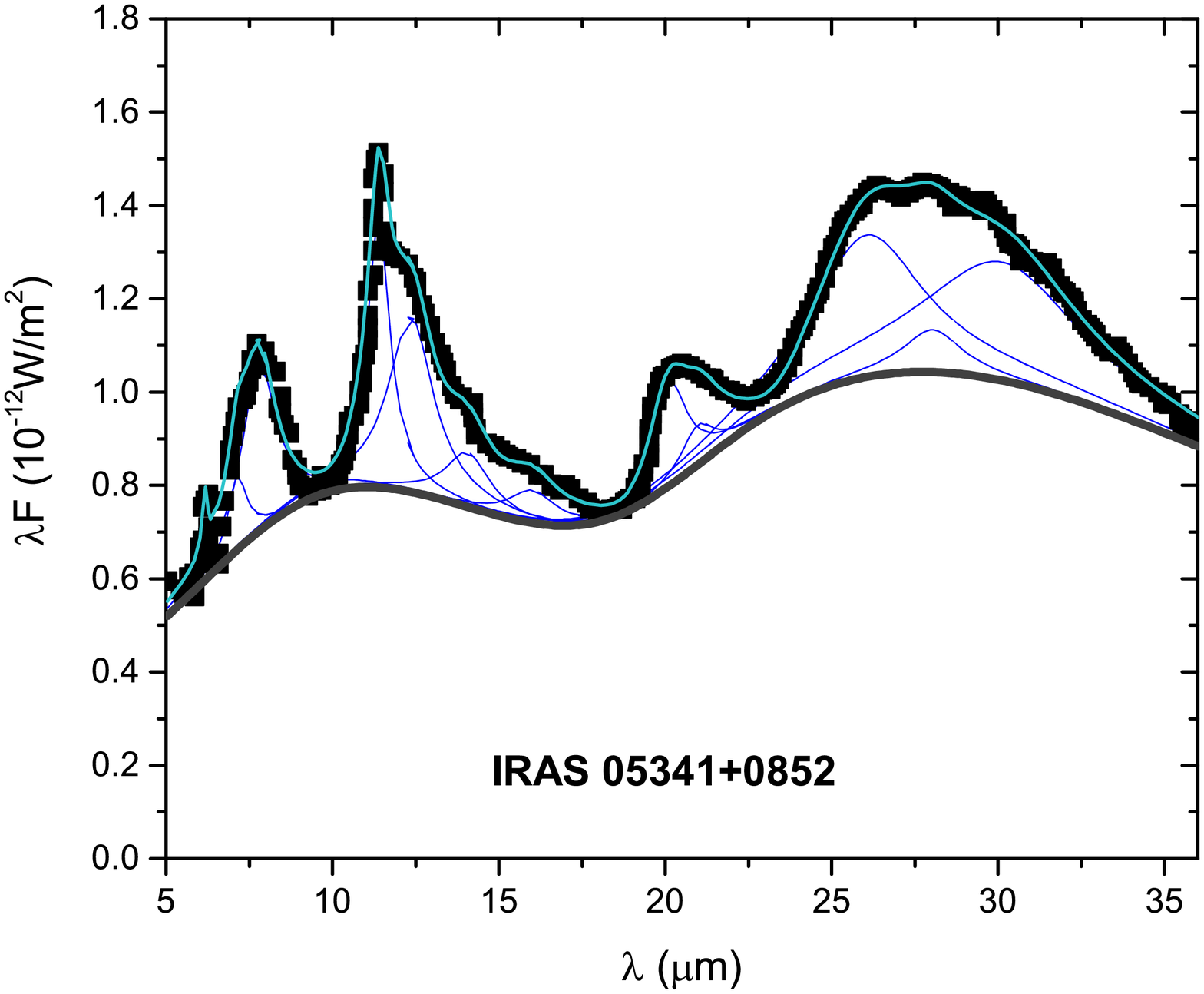} \\
    \includegraphics[width=.43\textwidth]{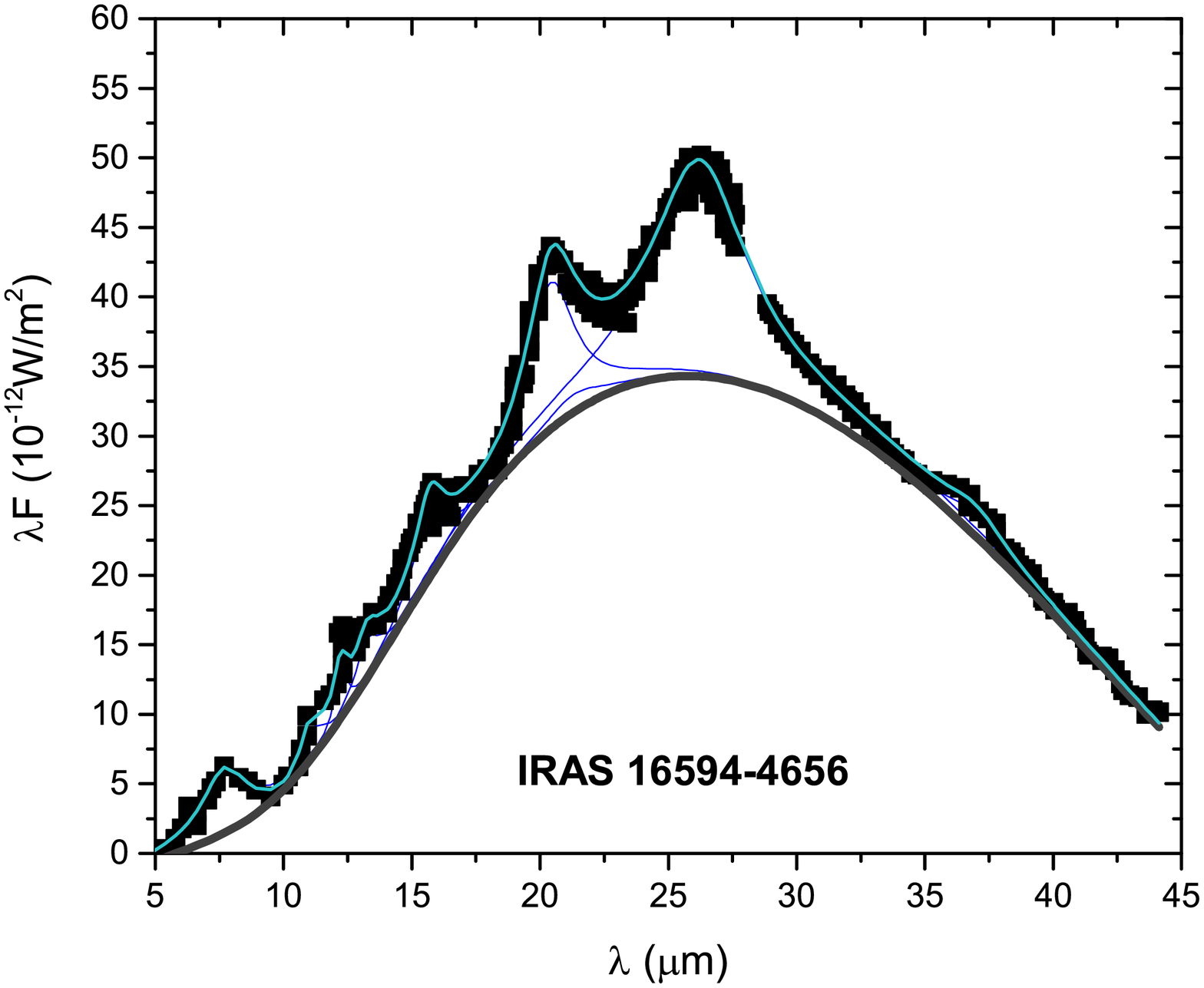} &
    \includegraphics[width=.43\textwidth]{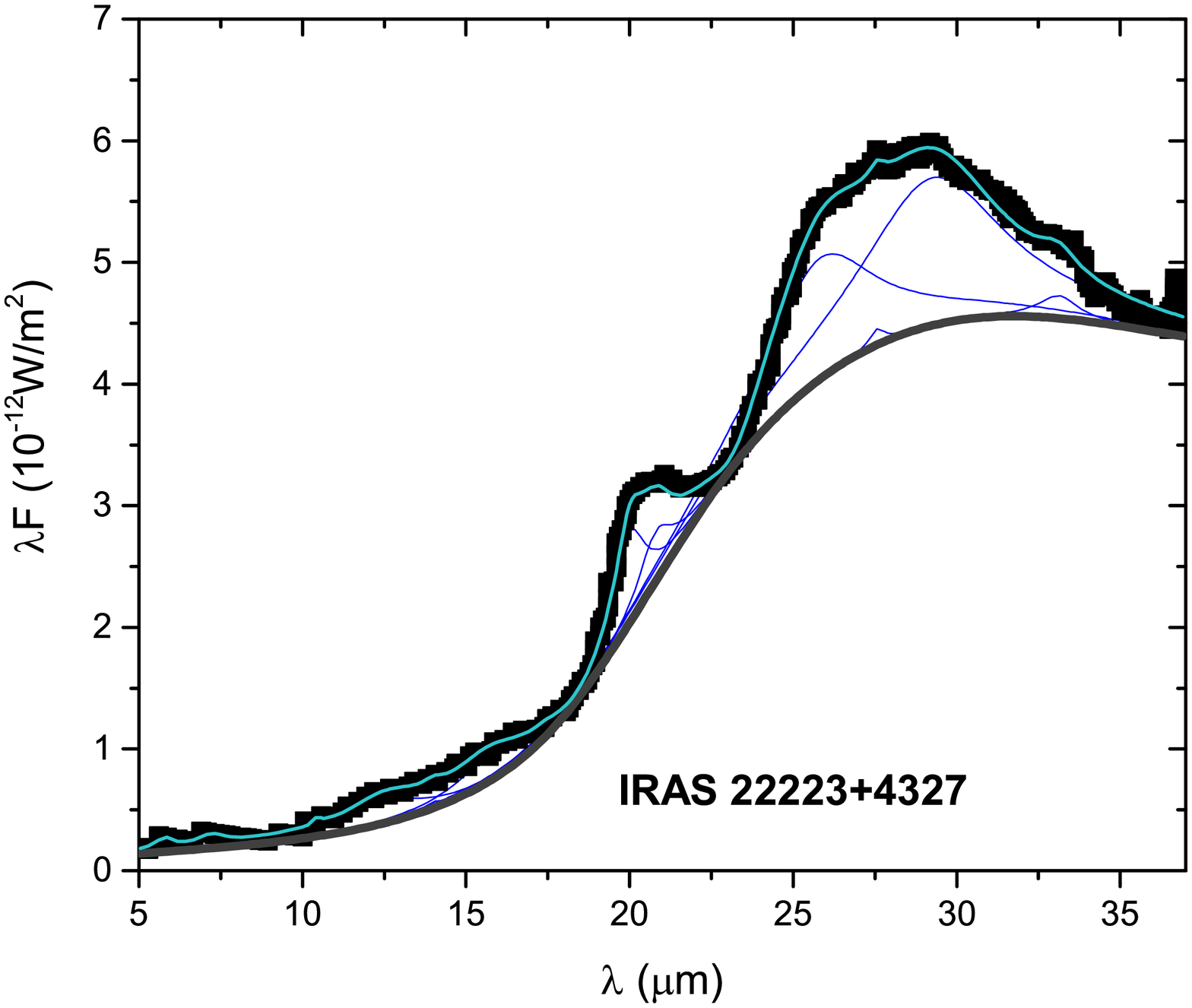}   \\
  \end{tabular}
\caption{\footnotesize
         \label{fig:spline_fit2}
         Same as Figure~\ref{fig:drudefit2}   
         but with the dust continuum 
         fitted with a spline function
         instead of two modified blackbodies. 
         }
\end{figure*}

\begin{figure*}
\centering
  \begin{tabular}{cc}
    \includegraphics[width=.43\textwidth]{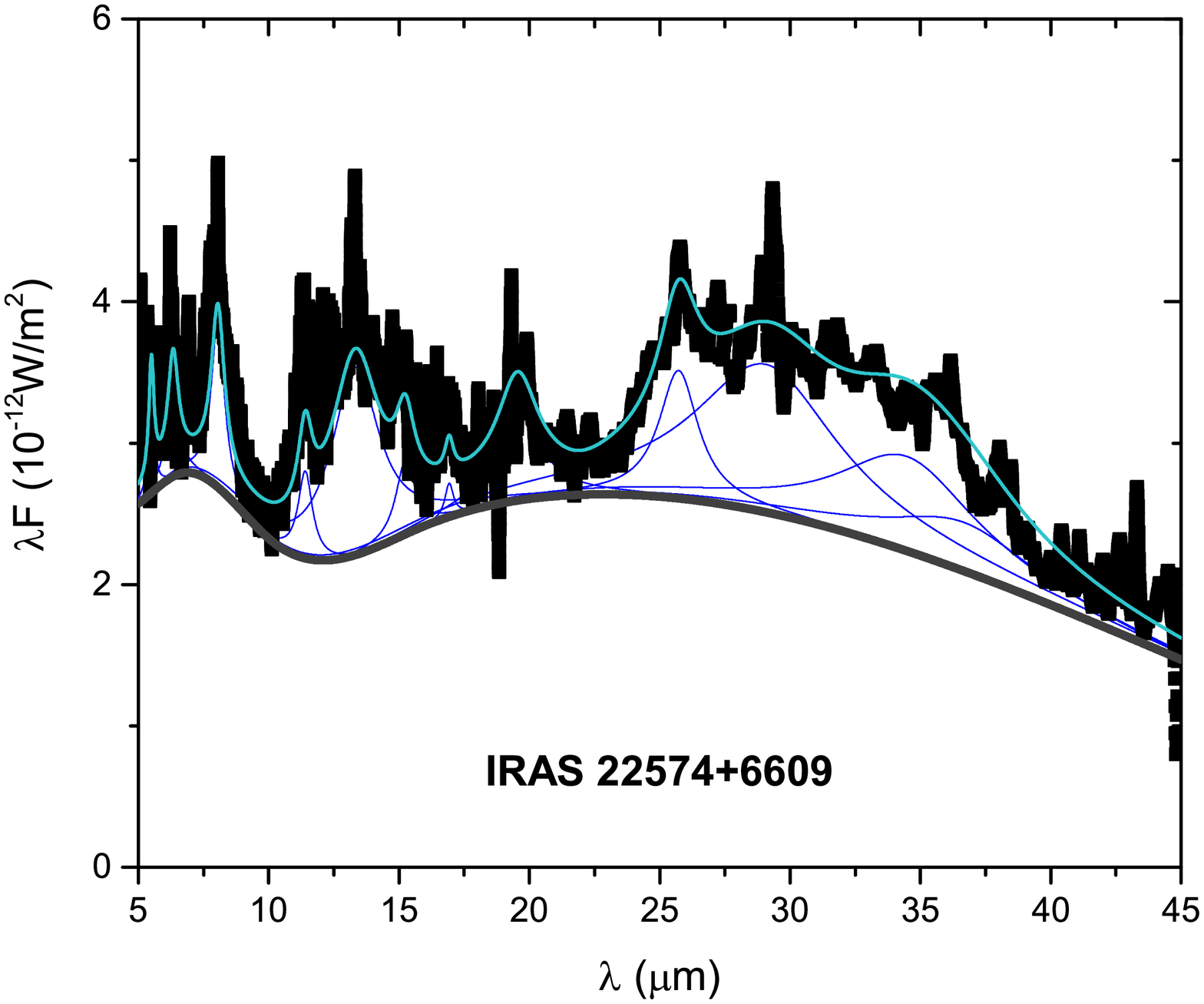} &
    \includegraphics[width=.43\textwidth]{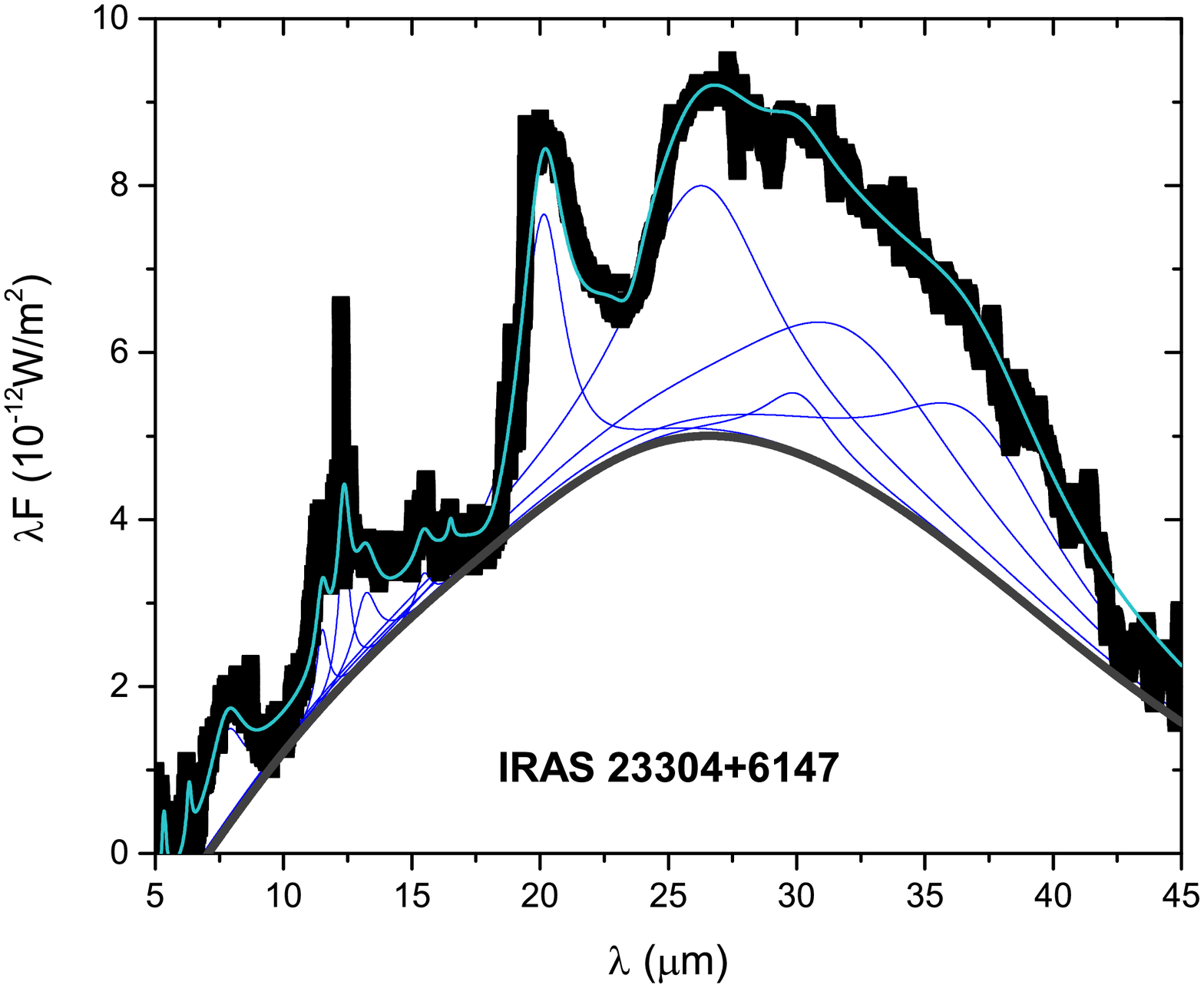}   \\   
  \end{tabular}
\caption{\footnotesize
         \label{fig:spline_fit3}
         Same as Figure~\ref{fig:drudefit3}   
         but with the dust continuum 
         fitted with a spline function
         instead of two modified blackbodies. 
         }
\end{figure*}

\section{PAHFIT Decomposition 
            Versus Spline Fitting
            \label{sec:continuum}}
\begin{figure*}
\centering
\vspace{-0.5cm}
\includegraphics[width=8.0cm]{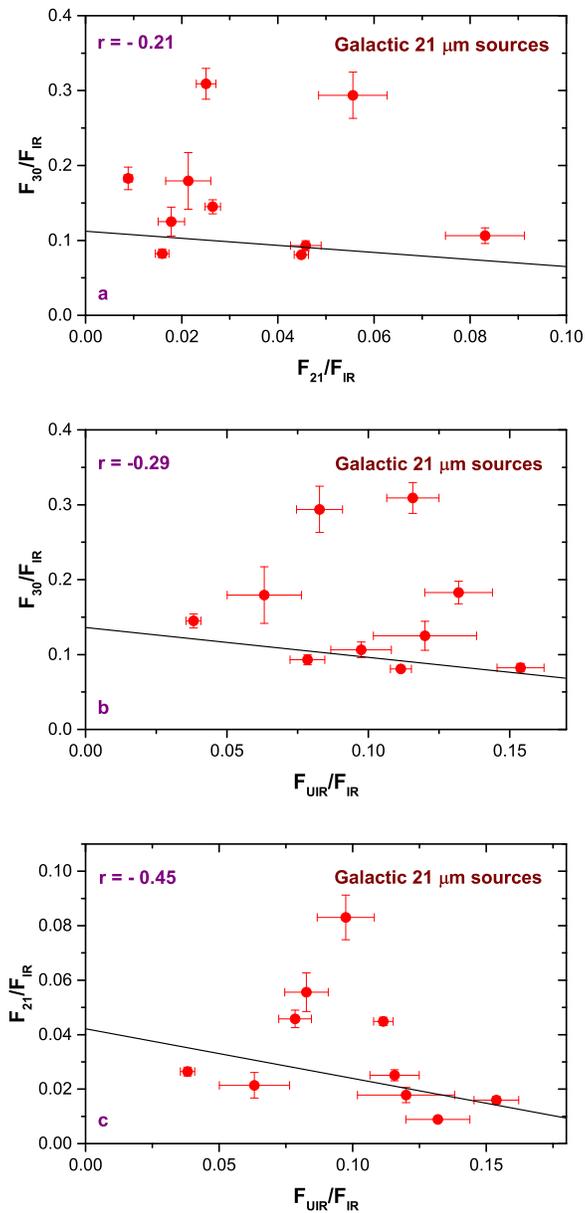}
\vspace{0.0cm}
\caption{\footnotesize
         \label{fig:2130uir_spline}
         Same as Figure~\ref{fig:2130uir_pahfit}
         but with the fluxes emitted in the 21$\mum$, 
         30$\mum$, and UIR features 
         derived from the spline continuum method
         (see \S\ref{sec:continuum}, and
          Figures~\ref{fig:spline_fit1},\ref{fig:spline_fit2}).
         }
\end{figure*}

\vspace{-1mm}
When we use the PAHFIT decomposition method to derive
the fluxes emitted from the 21$\mum$, 30$\mum$, 
and UIR features (see \S\ref{sec:fluxes}), we approximate 
the dust continuum with two modified blackbodies 
(see Figures~\ref{fig:drudefit1},\ref{fig:drudefit2}).
We admit that in dust spectral analysis the dust continuum
determination is often notoriously difficult. 
This is true for studying PAHs 
(e.g., see Uchida et al.\ 2000),
silicates 
(e.g., see Xie et al.\ 2014, Shao et al.\ 2014),
and many other dust species. 
For the 21$\mum$ sources discussed in this work,
what is the true dust continuum?
Apparently, the fluxes of the dust spectral features
are sensitive to the continuum determination. 
The key question here is to what extent 
the assumed continuum will affect 
the interrelations among these features 
examined in \S\ref{sec:2130uir}?
To examine this, instead of approximating
the dust continuum as two modified blackbodies,
we take several points 
at $\simali$6$\mum$, $\simali$10$\mum$, 
$\simali$18$\mum$, $\simali$24$\mum$, 
and $\simali$44$\mum$
to define an underlying continuum 
and fit the continuum with 
a cubic spline function 
(see Figures~\ref{fig:spline_fit1},\ref{fig:spline_fit2}).   
The 21$\mum$, 30$\mum$, and UIR features are again
fitted with Drude profiles. 
In Table~\ref{tab:SplineFlux} we list the fluxes 
emitted in the 21$\mum$, 30$\mum$ and UIR features 
derived from the spline continuum method,
as well as the dust IR flux integrated over 
the 5--45$\mum$ wavelength range.
They differ from that of the PAHFIT decomposition
method by $\simali$10\% for the 21$\mum$ and
UIR features, and by $\simali$30--60\% 
for the 30$\mum$ feature.
Figure~\ref{fig:2130uir_spline} shows
the interrelations of the fluxes
of the 21$\mum$, 30$\mum$ and UIR features
(normalized by the 5--45$\mum$ dust emission)
derived from the spline continuum method. 
Similar to that obtained from the PAHFIT 
decomposition method, we see no correlation
among these features.
Therefore, we conclude that although the true
dust continuum is not precisely known 
and the true fluxes emitted in the 21$\mum$, 
30$\mum$ and UIR features are sensitive to
the assumed continuum underneath the features,
these features are unlikely correlated 
as demonstrated by the two very different
continuum-determination methods.

\vspace{-6mm}
\section{Summary\label{sec:summary}}
\vspace{-1mm}
We explore the interrelations 
among the mysterious 21$\mum$,
30$\mum$ and UIR features 
of the  21$\mum$ sources. 
The principal results are the following:
\begin{enumerate}
\item The 21$\mum$ feature does not 
      correlate with the 30$\mum$ feature.
      This argues against
      the hypothesis of thiourea and aliphatic chains 
      attached to various carbonaceous structures as
      the carriers for both the 21$\mum$ feature
      and the 30$\mum$ feature.
\item The 21$\mum$ feature does not correlate with 
      the UIR features. This argues against large PAH 
      clusters as a possible carrier for the 21$\mum$ feature.
\item The 30$\mum$ feature and the UIR features
      are not correlated. 
      This does not support
      the speculation that the UIR carriers (e.g., PAHs) 
      may result from the decomposition or shattering of
      the 30$\mum$-feature carrier (e.g., HAC).
%
\end{enumerate}

\acknowledgments
We thank Z.~Ivezi\'c, T.~Ueta, 
K.~Volk, Y.X.~Xie, and K.~Zhang 
for helpful discussions.
We thank the anonymous referee
for his/her very helpful suggestions 
which substantially improved the quality
of this paper.
We are supported in part by
NSF AST-1109039, 
NNX13AE63G, 
NSFC\,11173019, 
and the University of Missouri Research Board.


\end{document}